\documentclass[10pt,journal,compsoc]{IEEEtran}
\ifCLASSOPTIONcompsoc
  \usepackage[nocompress]{cite}
\else
  \usepackage{cite}
\fi
\ifCLASSINFOpdf
  \usepackage[pdftex]{graphicx}
\else
\fi
\usepackage{algorithm}
\usepackage{algpseudocode} 
\usepackage{amsmath} 
\usepackage{amssymb} 

\begin{document}
%
\title{Enhancing Energy Efficiency in Scientific Workflows through CFD based PIVAEs}
%
%
%
%

\author{Ali Zahir, Ashiq Anjum, Mark Wilkinson and Jeyan Thiyagalingam 
\thanks{Manuscript received --; revised --. Ali Zahir is with the School of Computing and Mathematical Sciences, University of Leicester. (e-mail: ali.zahir4@gmail.com
).

Ashiq Anjum (Prof.) is with the School of Computing and Mathematical Sciences, University of Leicester. (e-mail: aa1180@leicester.ac.uk
).

Mark I. Wilkinson (Prof.) is with the Department of Physics and Astronomy, University of Leicester. (e-mail: miw6@leicester.ac.uk
).

Jeyan Thiyagalingam (Dr.) is with the Scientific Machine Learning Group, Rutherford Appleton Laboratory, Science and Technology Facilities Council (STFC–UKRI). (e-mail: t.jeyan@stfc.ac.uk
).  }}

%
%

\markboth{Journal of \LaTeX\ Class Files,~Vol.~14, No.~8, August~2023}%
{Shell \MakeLowercase{\textit{et al.}}: Bare Demo of IEEEtran.cls for Computer Society Journals}
%



\IEEEtitleabstractindextext{%
\begin{abstract}

The growing complexity and scale of scientific workflows in high performance computing (HPC) environments have led to significant challenges in managing energy consumption without compromising computational performance. Traditional scheduling strategies often fail to account for the complex interplay between thermal dynamics, workload diversity, and system scalability—leading to inefficient and unsustainable energy usage. This paper introduces a novel, scalable, and AI-assisted scheduling framework for optimizing energy consumption in high-performance computing (HPC) environments without compromising performance. Central to our approach is the integration of Computational Fluid Dynamics (CFD) with a Physics-Informed Variational Autoencoder (PIVAE), enabling the generation of physically realistic synthetic workload data that bridges the gap between thermodynamic behavior and scheduler decision-making in complex, multi-scale HPC environments. By categorizing workflows based on resource utilization profiles, we evaluate the impact of multiple scheduling strategies such as Locality Aware and Speculative Aware Scheduling on system performance and energy efficiency.These workflows—ranging from event reconstruction to anomaly detection—represent diverse computational intensities, demonstrating the adaptability of the proposed method across varying scientific domains. Our results show that modest reductions in CPU performance (e.g., to 15\%) can yield substantial energy savings (up to 10\%) with only minor turnaround time increases (approximately 5–6\%), identifying an optimal operational sweet spot. This work demonstrates how physics-informed generative modeling can enable adaptive, sustainable, and data-efficient scheduling decisions for next-generation HPC infrastructures. The proposed methodology offers a scalable path forward for sustainable data processing in modern distributed computing infrastructures.

\end{abstract}

}

\maketitle

\IEEEdisplaynontitleabstractindextext

%
\IEEEpeerreviewmaketitle

\IEEEraisesectionheading{\section{Introduction}\label{sec:introduction}}

\IEEEPARstart{L}ARGE-SCALE distributed computing systems, including high-performance computing (HPC) clusters and cloud platforms, form the backbone of modern scientific discovery in domains such as genomics, climate modeling, and high-energy physics, where complex, multi-stage computational workflows orchestrate data processing, simulation, and analysis tasks. These systems must process petabytes of data under stringent time constraints, while operating within limited power and cooling budgets. Energy inefficiency has therefore emerged as a first-order design constraint: studies estimate that CPU underutilization alone accounts for nearly 32\% of the total energy draw in HPC environments~\cite{versick2013power}, with additional waste stemming from thermal overheads, repetitive I/O, and imbalanced workload execution. Figure 1 illustrates the distribution of power consumption across typical computing system components, emphasizing the dominant role of CPU and cooling subsystems in overall energy draw. Despite continuous advancements in energy-efficient scheduling, existing approaches remain largely empirical and fail to integrate thermodynamic feedback into decision-making. Our work introduces a physics-informed, AI-driven approach that unifies data generation and scheduling under a single, physically realistic framework—addressing both the scale and sustainability challenges of next-generation HPC systems. As operational costs and carbon footprints escalate, energy-aware scheduling has become critical for both sustainability and performance.

Task scheduling plays a pivotal role in determining the trade-offs between turnaround time (TAT) and energy efficiency. Traditional schedulers such as First Come First Serve (FCFS) and Round Robin emphasize fairness and simplicity but do not account for heterogeneous hardware, thermal dynamics, or workload diversity~\cite{buyya2024energy}. More advanced heuristics such as Particle Swarm Optimization (PSO) ~\cite{dashti2016dynamic} or adaptive autoscaling ~\cite{iqbal2019adaptive,saxena2021proactive} have reduced energy costs through consolidation or prediction, yet they often incur high computational complexity and lack adaptability at scale. In contrast, we leverage physics-informed generative modeling to create synthetic, thermodynamically consistent data that empowers AI schedulers to learn optimal energy–performance trade-offs without exhaustive experimentation. Domain-specific strategies, including MapReduce-aware scheduling~\cite{yigitbasi2011energy}, GPU-aware placement using the energy-delay product (EDP)~\cite{zhang2013consolidating}, and DVFS-based offline scheduling~\cite{mei2017energy}, demonstrate effectiveness in narrow contexts but fail to generalize across diverse architectures. As HPC infrastructures continue to scale toward exascale and beyond, sustainability becomes a fundamental constraint—both in terms of energy efficiency and cooling overhead. Any viable scheduling approach must therefore operate effectively across large, heterogeneous environments while maintaining thermal and operational stability.

\begin{center}
\centering
\begin{figure}[]
\includegraphics[width=0.45\textwidth]{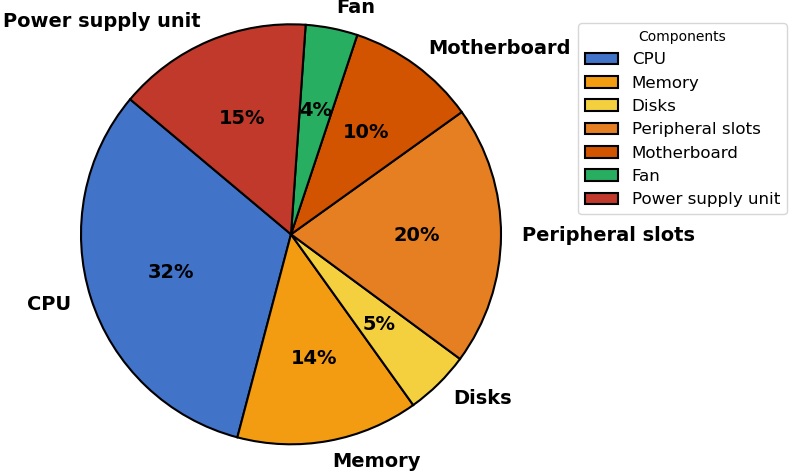}
\caption{Power consumption of a typical computing system components}
\end{figure}
\end{center}

Despite these advances, three fundamental gaps persist in energy-aware scheduling:

\begin{enumerate}
    \item \textbf{Thermal feedback is ignored.} Most existing methods model energy consumption only in terms of CPU frequency or utilization, neglecting the role of heat dissipation, airflow, and cooling inefficiencies. In HPC systems, thermal hotspots directly affect processor reliability and power draw, yet scheduling decisions remain thermally blind.

    \item \textbf{Synthetic data lacks physical realism.} Generative models such as Variational Autoencoders (VAEs)~\cite{gupta2020vae,pinheiro2021variational} have been applied to overcome data scarcity, but standard VAEs produce synthetic workload traces that violate physical constraints. Without grounding in thermodynamics, these datasets lead to scheduling policies that are infeasible under real-world conditions.

    \item \textbf{Over-specialization to hardware.} Several approaches target specific ecosystems e.g., Hadoop clusters~\cite{yigitbasi2011energy}, GPU-centric schedulers~\cite{zhang2013consolidating}, or CPU-GPU power capping~\cite{venkataswamy2022rare}. While effective within their domains, such methods lack generality and are difficult to extend across heterogeneous infrastructures.
\end{enumerate}

These gaps create a misalignment between theoretical scheduling models and practical HPC deployment, limiting the impact of current energy-aware strategies. In particular, the absence of thermodynamically grounded data hinders the design of adaptive schedulers capable of balancing energy efficiency with performance.

To address these challenges, we propose a \textit{Physics Informed Variational Autoencoder (PIVAE)} coupled with \textit{Computational Fluid Dynamics (CFD)} modeling. CFD provides fine-grained simulations of heat propagation, cooling efficiency, and dynamic power dissipation, while the PIVAE ensures that the synthetic data generated for scheduler training respects thermodynamic constraints. This integration enables the adaptive exploration of operational scenarios that are both statistically valid and physically realistic. 

We evaluate the CFD-PIVAE system across five representative scientific workflows—Event Reconstruction (WF-1), Particle Trajectory Identification (WF-2), Collision Point Detection (WF-3), Pattern Recognition (WF-4), and Anomaly Detection (WF-5)—using multiple schedulers, including First-Come, First-Served (FCFS), Locality-Aware Scheduling (LAS), LAS with Prefetching (LASP), Speculative-Aware Scheduling (SAS)~\cite{zahir2025sas}, LYNX, and the recent OM-FNN approach. Results show that CFD-PIVAE guided scheduling achieves up to 10\% energy savings with only a 5-6\% increase in TAT, outperforming conventional methods. Importantly, while LYNX achieves TAT reductions at the cost of higher energy consumption and  OM-FNN achieves less energy consumption at the cost of higher TAT, our approach consistently delivers balanced trade-offs.

The contributions of this paper are as follows:
\begin{itemize}
    \item A novel CFD-informed PIVAE architecture that generates thermodynamically realistic synthetic datasets for energy-aware scheduling.
    \item A workflow-specific optimization system that balances CPU utilization, memory access, and thermal constraints to reduce energy consumption without degrading Turnaround time.
    \item Empirical validation on a Proxmox-managed HPC cluster, demonstrating scalable energy savings compared with FCFS, LAS, SAS, LYNX, and OM-FNN.
\end{itemize}

The remainder of this paper is organized as follows: Section~2 reviews related work on energy-aware scheduling. Section~3 presents the proposed CFD-PIVAE architecture. Section~4 details the experimental setup and workflow characterization. Section~5 discusses results across multiple schedulers, and Section~6 concludes with future directions.

\section{Literature Review}

Energy aware scheduling in distributed computing has advanced from early fairness-oriented methods like Round Robin and FCFS~\cite{buyya2024energy} to optimization-driven approaches such as PSO-based virtual machines placement~\cite{dashti2016dynamic}, which improved energy use but suffered from scalability issues. Machine learning further refined scheduling: Gupta et al.~\cite{gupta2018efficient} employed VAEs to model energy patterns, while Iqbal et al.~\cite{iqbal2019adaptive} introduced adaptive multi-cloud algorithms. However, these often overlooked hardware constraints such as CPU-GPU interactions and thermal dynamics~\cite{mei2017energy}. Zahir et al.~\cite{zahir2025sas} extended scheduling with locality awareness, prefetching, and speculative execution, but without addressing energy footprints. Xie et al.~\cite{xie2017energy} present energy-efficient scheduling algorithms based on DVFS for real-time parallel applications on heterogeneous distributed systems, achieving substantial energy savings under deadline constraints; however, their approach relies on static analytical models and does not account for thermal dynamics or uncertainty in workload behavior. In contrast, our work addresses these limitations by integrating CFD-informed thermal constraints and probabilistic modeling via PI-VAE to enable energy-aware scheduling under realistic, data-driven, and uncertainty-aware operating conditions.

Recent studies have explored predictive and modeling-based methods. Saxena et al.~\cite{saxena2021proactive} proposed multi-resource prediction for autoscaling, Aldossary et al.~\cite{aldossary2018performance} used ARIMA to trade off cost and energy, and Dashti et al.~\cite{dashti2016dynamic} applied PSO for consolidation. Other domain-specific solutions include MapReduce profiling~\cite{yigitbasi2011energy}, EDP-based GPU job placement~\cite{zhang2013consolidating}, DVFS task scheduling~\cite{mei2017energy}, and ML-driven CPU-GPU co-scheduling with power capping~\cite{venkataswamy2022rare}. Swarm intelligence methods have also been reviewed for industrial contexts~\cite{gao2020review}. However, despite these advances, most existing approaches remain limited by their reliance on statistical or heuristic models that ignore underlying physical processes such as heat transfer and cooling dynamics. As a result, their predictions often lack thermodynamic realism and fail to generalize across heterogeneous HPC environments, motivating the need for a physics-informed, scalable framework such as the proposed CFD-PIVAE system.

At the infrastructure level, virtualization and consolidation remain central~\cite{shuja2014survey,alam2017study}, and learning-centric cloud designs reduce carbon footprints~\cite{buyya2024energy}, but most approaches neglect dynamic workload and thermal variability. Few works integrate thermal feedback into scheduling, despite its impact on reliability and energy efficiency. CFD offers fine-grained modeling of heat transfer, yet lacks adaptability. To address this, Physics-Informed VAEs (PIVAEs) combine CFD with generative modeling to create thermodynamically consistent synthetic data, enabling realistic exploration of scheduling trade-offs between turnaround time and energy consumption.

Synthetic data generation has been applied using GANs, Monte Carlo methods~\cite{figueira2022survey}, ARMs~\cite{viana2024synthetic}, RBMs~\cite{mnih2012conditional}, and VAEs~\cite{pinheiro2021variational}. While VAEs stabilize training and capture uncertainty, most models lack thermodynamic realism. Integrating physics into generative systems~\cite{saxena2021proactive,zhang2013consolidating} provides better domain fidelity. However, gaps remain: reliance on simplified energy models~\cite{yigitbasi2011energy}, hardware-specific constraints~\cite{mei2017energy}, and synthetic data divorced from physical laws~\cite{aldossary2018performance}. This motivates our proposed CFD-guided PIVAE approach, which enforces thermodynamic plausibility while supporting adaptive, energy-efficient scheduling.

\section{Proposed Solution}

We propose a physics-informed energy-aware scheduling system for High Performance Computing (HPC) workflows that integrates \textit{Computational Fluid Dynamics (CFD)} with a \textit{Physics Informed Variational Autoencoder (PIVAE)}. Unlike conventional data-driven methods that rely solely on statistical modeling, this system embeds physical constraints such as thermal dynamics and power dissipation directly into scheduling decisions, ensuring that generated operational configurations remain both realistic and deployable in practice.

The system operates in three stages. First, execution traces are collected from representative workflows under varying CPU frequencies and scheduling strategies. These traces include execution time, energy consumption, and resource utilization metrics, forming the training dataset for the PIVAE. Second, CFD simulations are employed to capture heat dissipation, airflow behavior, and temperature distribution across HPC components under different workload scenarios. Instead of treating CFD outputs as post-analysis artifacts, their temperature-aware power profiles are integrated into the PIVAE’s latent space, constraining the generative process so that synthetic data reflects both statistical workload behavior and thermodynamic feasibility. Finally, the trained PIVAE produces synthetic datasets that extend the configuration space beyond what is feasible through brute-force experimentation, enabling the discovery of operational scenarios that balance turnaround time (TAT) and energy consumption while respecting physical limits.

To connect these modeling capabilities with real scheduling decisions, the system incorporates a scheduling decision module that evaluates workflow requirements and recommends suitable task allocation strategies. We consider six representative schedulers:First Come First Serve (FCFS), which ensures fairness but often neglects efficiency; Locality-Aware Scheduling (LAS), which prioritizes data placement to reduce I/O delays; LAS with Prefetching (LASP), which anticipates overlapping data access patterns; LYNX, which extends locality optimization through predictive prefetching; Speculative-Aware Scheduling (SAS), which uses predictive modeling to enable speculative task execution; and OM-FNN, a machine learning–based approach that leverages optimized feedforward neural networks to predict energy consumption patterns and guide scheduling choices. Together, these schedulers span heuristic-based, predictive, speculative, and learning-driven philosophies, providing a diverse set of strategies through which the system can generate and evaluate physically consistent scheduling scenarios.

By constraining synthetic data generation with CFD-derived thermal models and applying it across multiple scheduling paradigms, the proposed system enables adaptive scheduling decisions that account for workload diversity, hardware heterogeneity, and thermodynamic constraints in HPC environments. This design unifies data-driven flexibility with physics-based fidelity, providing a principled pathway for energy-aware scheduling that remains realistic, scalable, and broadly applicable.

Figure 2 illustrates the overall architecture to identify optimal operational configurations using PIVAE. The VAE utilizes the time and energy consumption data from various scheduling algorithms as input (real data) to generate similar synthetic data, which aids in determining the sweet spot for operational configurations.

\begin{figure}[h!]
\centering
\includegraphics[width=0.5\textwidth]{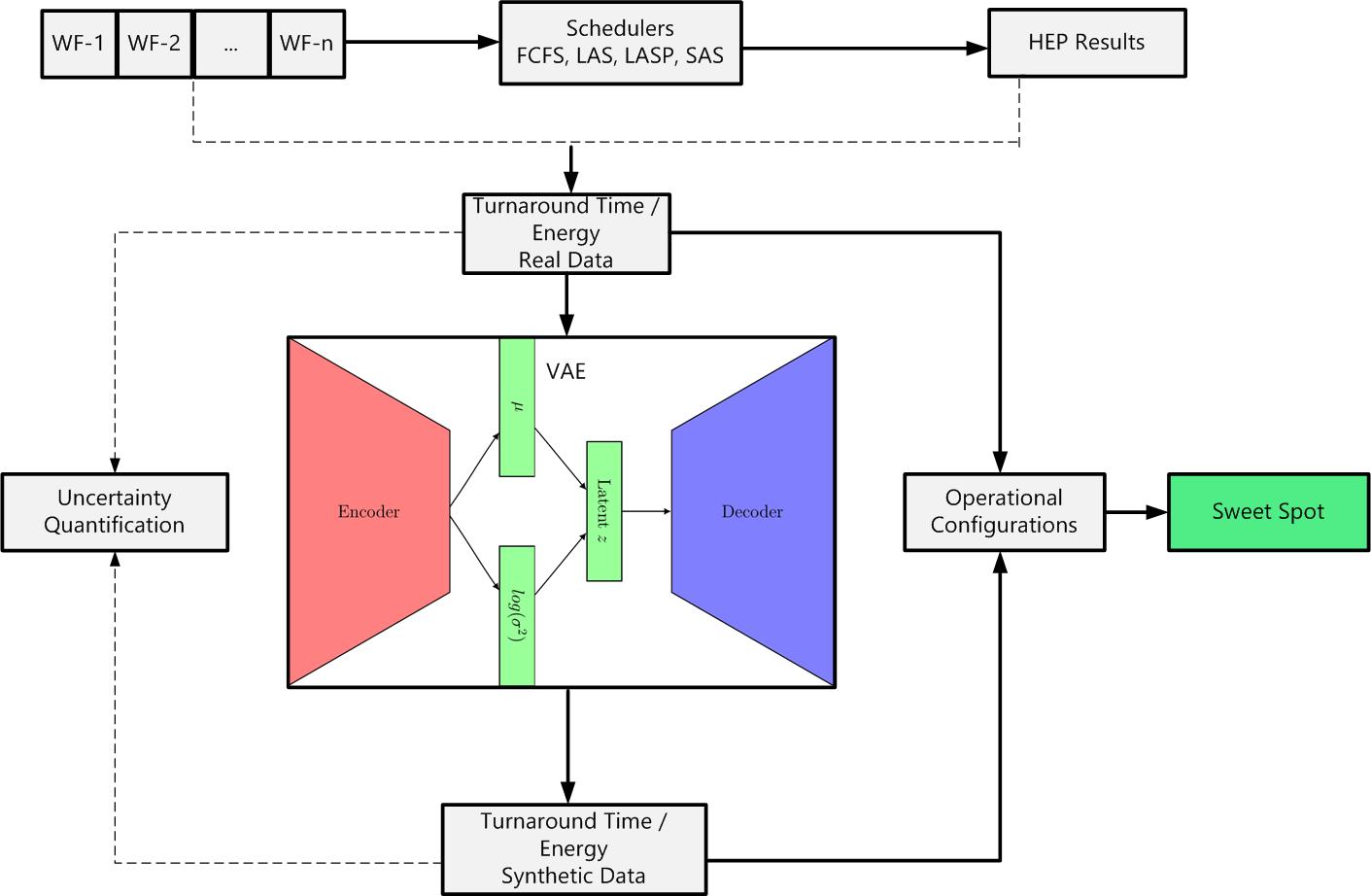}
\caption{CFD-PIVAE Architecture Diagram}
\end{figure}

\subsection{Energy Consumption by Using Computational Fluid Dynamics}

We use Computational Fluid Dynamics (CFD) to account for the physical principles of heat transfer, thermal conduction, and cooling efficiency, which directly influence processor energy consumption. The CFD provides temperature and heat dissipation profiles under different workload scenarios. These outputs are then integrated into our mathematical energy model, which combines processor power dissipation, memory and I/O activity, and cooling mechanisms into a unified representation of total energy usage. In this way, CFD contributes to the temperature dependent components of the model, ensuring that energy consumption is estimated under conditions that reflect realistic thermal behavior in HPC systems.

To validate and calibrate the CFD simulations, empirical power and thermal data were collected from the underlying high-performance computing (HPC) cluster during workflow execution. Power measurements were obtained using Intel Running Average Power Limit (RAPL) interfaces for per-core energy consumption and IPMI/BMC sensors for system-level power draw, sampled at 100 ms intervals. Thermal data—including CPU core, memory DIMM, and inlet/outlet temperatures—were continuously monitored via LM-Sensors and logged through a Prometheus-based telemetry stack. These real-time readings were used both to parameterize CFD boundary conditions and to cross-verify simulated temperature and heat-flow profiles, ensuring that the modeled results corresponded closely to observed hardware behavior.

\subsubsection{Total Energy Model}
The total energy consumption, \(E_{\text{total}}\), during the execution of a workflow over time \(t\), is given by:
\begin{equation}
    E_{\text{total}} = \int_0^t P_{\text{total}}(t) \, dt,
\end{equation}
where \(P_{\text{total}}\) represents the total power consumption of the system, modeled as:
\begin{equation}
    P_{\text{total}} = P_{\text{processor}} + P_{\text{memory}} + P_{\text{I/O}} + P_{\text{cooling}}.
\end{equation}

\subsubsection{Processor Power Consumption}
The power consumption of the processor, \(P_{\text{processor}}\), is the sum of dynamic and static power:
\begin{equation}
    P_{\text{processor}} = P_{\text{dynamic}} + P_{\text{static}}.
\end{equation}

\begin{itemize}
    \item Dynamic Power
\end{itemize}
Dynamic power depends on the processor's switching activity, supply voltage, and operating frequency:
\begin{equation}
    P_{\text{dynamic}} = C \cdot V^2 \cdot f,
\end{equation}
where \(C\) is the effective switching capacitance, \(V\) is the supply voltage, and \(f\) is the clock frequency. Memory and I/O intensive workflows typically involve reduced processor utilization, resulting in lower \(P_{\text{dynamic}}\).

\begin{itemize}
    \item Static Power
\end{itemize}
Static power arises from leakage currents and is temperature-dependent:
\begin{equation}
    P_{\text{static}} = I_{\text{leakage}}(T) \cdot V,
\end{equation}
where \(I_{\text{leakage}}(T)\) is the leakage current at a given temperature \(T\), and \(V\) is the supply voltage. The temperature \(T\) is obtained from CFD simulations of heat transfer within the processor, modeled by the heat equation \cite{bergman2011fundamentals}\cite{ZHONG2023115664}:
\begin{equation}
    \rho c_p \frac{\partial T}{\partial t} + \rho c_p \mathbf{v} \cdot \nabla T = k \nabla^2 T + Q 
\end{equation}
Here, \(\rho\) is the density of the material, which corresponds to the processor and the heat sink. \(c_p\) is the specific heat capacity, \(\mathbf{v}\) is the velocity field of airflow (if forced convection is present), \(k\) is the thermal conductivity and \(Q\) is the heat source derived from \(P_{\text{dynamic}}\).

\subsubsection{Memory Power Consumption}

Memory power consumption, \(P_{\text{memory}}\), is influenced by dynamic activity, refresh cycles, and standby leakage. Dynamic power arises from read/write operations:
\begin{equation}
    P_{\text{memory-dynamic}} = N_{\text{reads}} \cdot E_{\text{read}} + N_{\text{writes}} \cdot E_{\text{write}},
\end{equation}
where workload intensity determines the number of accesses. DRAM refresh power is modeled as:
\begin{equation}
    P_{\text{refresh}} = f_{\text{refresh}} \cdot E_{\text{refresh}},
\end{equation}
which grows with temperature and is informed by CFD-based thermal profiles. Idle power stems from leakage currents:
\begin{equation}
    P_{\text{memory-idle}} = I_{\text{leakage}}(T) \cdot V,
\end{equation}
where leakage increases at higher operating temperatures. 

The total memory power is thus:
\begin{equation}
    P_{\text{memory}} = P_{\text{memory-dynamic}} + P_{\text{refresh}} + P_{\text{memory-idle}},
\end{equation}
capturing workload activity, refresh overhead, and thermal effects in HPC environments.

\subsubsection{I/O Power Consumption}
The power consumed by I/O operations, \(P_{\text{I/O}}\), includes active and idle components:
\begin{equation}
    P_{\text{I/O}} = P_{\text{I/O-active}} + P_{\text{I/O-idle}}.
\end{equation}

\begin{itemize}
    \item Active Power:
\end{itemize}

\begin{equation}
    P_{\text{I/O-active}} = R_{\text{I/O}} \cdot E_{\text{I/O}},
\end{equation}
where \(R_{\text{I/O}}\) is the data transfer rate, and \(E_{\text{I/O}}\) is the energy per byte transferred.

\begin{itemize}
    \item Idle Power
\end{itemize}

\begin{equation}
    P_{\text{I/O-idle}} = P_{\text{idle}} \cdot (t_{\text{total}} - t_{\text{active}}).
\end{equation}

\subsubsection{Cooling Power Consumption}
Cooling power, \(P_{\text{cooling}}\), accounts for the energy required to dissipate the heat generated by the processor and other components. Using CFD-derived heat removal rates, \(Q_{\text{removed}}\), the cooling power is given by:
\begin{equation}
    P_{\text{cooling}} = \frac{Q_{\text{removed}}}{\eta_{\text{cooling}}},
\end{equation}
where \(\eta_{\text{cooling}}\) is the efficiency of the cooling system.

\subsubsection{Final Energy Model}
Combining all components, the total energy consumption for memory- and I/O-intensive workflows is expressed as:

\begin{equation}
\begin{split}
    E_{\text{total}} = \int_0^t \Big( P_{\text{dynamic}} + P_{\text{static}} + P_{\text{memory-active}} + P_{\text{memory-idle}} \\
    + P_{\text{I/O-active}} + P_{\text{I/O-idle}} + \frac{Q_{\text{removed}}}{\eta_{\text{cooling}}} \Big) dt.
\end{split}
\end{equation}

This model provides a system for analyzing energy consumption in memory and I/O intensive workflows, accounting for processor dynamics, memory and I/O operations, and thermal effects derived from CFD simulations. The priority of the workflows can be used to determine the order in which the workflows are executed, and hence, the order in which the energy consumption is calculated.

When calculating energy consumption for CPU usage in data analysis workflows, the focus is on accurately measuring the energy spent on CPU and I/O operations. Given that CPU energy consumption accounts for a significant proportion, approximately 37\%, of the total energy usage as highlighted in earlier sections, this focus enables a precise assessment of energy efficiency across various computational workflows. By employing the calculations for CPU energy consumption (E\_CPU), alongside considerations for memory (E\_mem), storage (E\_storage), and network (E\_network) energy expenditures, the primary objective was to aggregate the energy consumption for each workflow.

The scheduling strategies examined within the research context primarily focus on minimizing turnaround times for computational workflows. However, it's crucial to recognize that these strategies have varying impacts on resource utilization and energy consumption. A scheduling algorithm that excels in reducing turnaround time does not automatically equate to being energy efficient. For instance, efficient scheduling algorithms like speculation or locality aware, while adept at cutting down on turnaround time, tend to be more demanding in terms of computational resources. This higher computational demand invariably leads to increased energy consumption, making these processes more power intensive. In essence, the pursuit of minimized turnaround times through these scheduling strategies must be balanced with considerations for their energy and resource implications, highlighting the complex trade offs involved in optimizing computing workflows.

\subsection{VAEs for Energy Consumption Modeling}
While real-world workflow traces and energy measurements are inherently limited in scope, this challenge directly motivates our approach. The proposed CFD-PIVAE framework addresses data scarcity by generating physically realistic synthetic samples that emulate diverse operational scenarios. This enables the exploration of energy–performance trade-offs beyond what is feasible through conventional empirical experimentation. We have used a Variational Autoencoder (VAE) to generate synthetic data that represents the connection between energy usage and turnaround time. This data helps predict the best configurations that balances energy efficiency and computational performance in the best possible manner.

The dataset contained important parameters such as the number of workflows (N), input data requirements (IR), system usage, turnaround time (TAT), and energy consumption (E) linked to each workflow. Standardizing the dataset ensured consistency among features. We then created the VAE structure by defining the encoder and decoder networks, and selecting the optimal number of layers and activation functions. The loss function, which includes reconstruction loss, was established to direct model optimisation. The VAE was trained by dividing the dataset into training and validation sets. Backpropagation and gradient descent optimisation algorithms were used to iteratively update model parameters in order to minimise the total loss. The optimal configuration for the VAE was determined by tuning hyperparameters and evaluating on the validation set. 

After training, the VAE model made synthetic data from the learned latent space that showed different operational situations and how they affected the turnaround time.

The synthesised data helped analyse the complex connections among operational configuration, turnaround time, and energy consumption in the analysis workflows.
\subsubsection{Data Preprocessing and Encoder Design}

This section describes the transformation of raw scheduling and energy data into latent representations used by the proposed CFD-PIVAE framework. The process integrates data preprocessing with a physics-informed encoder to ensure that both statistical and thermodynamic characteristics are preserved.

\textbf{Data Preparation:}
Let $X$ be the input dataset containing the following features:

\begin{itemize}
  \item Workflow (WF)
  \item Number of tasks (TASKS)
  \item Processing time per task
  \item CPU frequency usage
  \item Turnaround time (TAT) in milliseconds
  \item Power consumption by CPU
  \item Energy consumption in Joules
  \item Energy consumption in kWh
\end{itemize}

Given a dataset $\mathcal{D}$ consisting of tuples $(\mathbf{x}_i, \mathbf{y}_i, \mathbf{n}_i, w_i, t_i, p_i, e_i, k_i)$, where $\mathbf{x}_i$ represents the scheduling technique, $\mathbf{y}_i$ the workflow, $\mathbf{n}_i$ the number of tasks, $t_i$ the processing time per task, $w_i$ the CPU frequency usage, $p_i$ the CPU power consumption, $e_i$ the energy consumption in Joules, and $k_i$ the energy consumption in kWh for each workflow instance $i$.

\textbf{Preprocessing Steps:}
\begin{enumerate}
    \item \textbf{Categorical Encoding:} Categorical variables $\mathbf{x}_i$ and $\mathbf{y}_i$ are one-shot encoded. This mapping $f: \mathcal{C} \rightarrow \mathbb{R}^n$ produces a binary vector $f(c)$ for each category $c \in \mathcal{C}$, where $n$ is the number of unique categories.
    
    \item \textbf{Numerical Feature Scaling:} Continuous variables $\mathbf{n}_i, w_i, t_i, p_i, e_i, k_i$ are scaled to $[0,1]$ using:
    \begin{equation}
        g(v) = \frac{v - \min(\mathbf{v})}{\max(\mathbf{v}) - \min(\mathbf{v})}
    \end{equation}
    where $\mathbf{v}$ represents all instances of a given feature in $\mathcal{D}$.
    
    \item \textbf{Data Integration:} The encoded categorical and scaled numerical features are concatenated to form the input matrix $\mathbf{X} \in \mathbb{R}^{m \times k}$, where $m$ is the number of samples and $k$ the total number of processed features. Missing values are imputed via mean substitution.
    
    \item \textbf{Dataset Splitting:} $\mathbf{X}$ is partitioned into training ($\mathbf{X}_{train}$) and validation ($\mathbf{X}_{val}$) sets.
\end{enumerate}

These preprocessing steps ensure that scheduling configurations, CPU frequencies, and power metrics are consistently normalized for input to the encoder network.

\textbf{Encoder Design:}
The encoder component in the proposed CFD-PIVAE differs from conventional VAEs by incorporating both physics-informed and scheduler-aware features. In addition to conventional numerical parameters (e.g., turnaround time, energy consumption, CPU frequency), the encoder integrates thermodynamic indicators obtained from CFD simulations—such as temperature gradients and heat dissipation coefficients—alongside categorical scheduler identifiers (e.g., FCFS, LAS, LASP, SAS, LYNX, OM-FNN). This composite representation enables the model to learn latent embeddings that capture statistical relationships as well as physical dependencies in workflow energy behavior.

The encoder computes the mean and variance of the latent distribution as follows:
\begin{equation}
\mu(x), \log \sigma^2(x) = \text{EncoderNetwork}(x)
\end{equation}
\begin{equation}
z = \mu(x) + \exp\left(\frac{1}{2} \log \sigma^2(x)\right) \cdot \epsilon, \quad \epsilon \sim \mathcal{N}(0, I)
\end{equation}

where $x$ denotes the input vector after preprocessing, $\mu(x)$ and $\sigma(x)$ represent the parameters of the latent distribution, and $z$ is the sampled latent variable.

During training, the encoder aligns latent representations with CFD-derived thermal distributions, ensuring that encoded features maintain thermodynamic realism. This allows the VAE to generate synthetic workload configurations that are both statistically valid and physically consistent—enabling data-driven exploration of energy–performance trade-offs across diverse scheduling strategies.

\subsubsection{Decoder, Loss Function, and Validation Process}

The decoder and loss formulation together define how the CFD-PIVAE reconstructs realistic energy–performance profiles while enforcing thermodynamic consistency. After the encoder produces a latent variable $z$, the decoder reconstructs the input features and estimates corresponding energy and turnaround-time relationships.

\textbf{Decoder:}
The decoder network reconstructs the input data from the latent representation by minimizing the difference between the original and generated samples while respecting energy-efficiency constraints. This reconstruction enables the exploration of alternative operational configurations that balance energy consumption and performance across various workflows and scheduling techniques:
\begin{equation}
\hat{x} = \text{DecoderNetwork}(z)
\end{equation}
where $\hat{x}$ denotes the reconstructed feature vector derived from the latent variable $z$.

\textbf{Loss Function:}
To ensure physical validity and stable latent representations, the total loss integrates three complementary objectives:
\begin{itemize}
    \item \textbf{Reconstruction Loss ($\mathcal{L}_{rec}$):} Preserves operational characteristics of the original data.
    \begin{equation}
        \mathcal{L}_{rec} = || x - \hat{x} ||^2
    \end{equation}

    \item \textbf{Kullback–Leibler Divergence ($\mathcal{L}_{KL}$):} Regularizes the latent space for smooth and continuous sampling.
    \begin{equation}
        \mathcal{L}_{KL} = D_{KL}\big(q(z|x) \,||\, p(z)\big)
    \end{equation}
     Here, $q(z|x)$ represents the encoder’s learned posterior distribution over latent variables, and $p(z)$ denotes the prior distribution, typically modeled as a standard normal.

    \item \textbf{CFD-Based Energy Constraint ($\mathcal{L}_{CFD}$):} Adds a physics-informed regularization term derived from CFD-simulated power dissipation.
    \begin{equation}
        \mathcal{L}_{CFD} = \sum_{i=1}^{N} \big| E_{CFD}(x_i) - \hat{E}(x_i) \big|
    \end{equation}
    Here, $E_{CFD}(x_i)$ is the energy predicted by CFD simulations for configuration $x_i$, and $\hat{E}(x_i)$ is the corresponding VAE-estimated energy.
\end{itemize}

The total composite loss is formulated as:
\begin{equation}
    \mathcal{L} = \mathcal{L}_{rec} + \beta \mathcal{L}_{KL} + \gamma \mathcal{L}_{CFD}
\end{equation}
where $\beta$ and $\gamma$ control the trade-off between latent-space regularization and CFD-based physical realism.

\textbf{Validation Process:}
After training, the generated synthetic data undergoes multi-stage validation to ensure thermodynamic plausibility and operational feasibility:

\begin{itemize}
    \item \textbf{Data Generation:} 
    \begin{equation}
        Y = VAE(Data), \quad
        Y' = \{\, y_i \in Y \mid D(y_i) = \text{True} \,\}
    \end{equation}
    $Y'$ represents samples that satisfy CFD-based energy constraints.

    \item \textbf{Constraint Verification:}
    \begin{equation}
        V(Y, F) = \{\, y_i \in Y \mid \forall f \in F,\, T(y_i, f) = \text{True} \,\}
    \end{equation}
    Each generated sample must meet predefined power and thermal thresholds derived from real-system measurements.

    \item \textbf{Cross-Validation and Sensitivity Analysis:}
    \begin{equation}
        C(Y, S) = \{\, (y_i, s_j) \mid y_i \in V(Y, F),\, s_j \in S \,\}
    \end{equation}
    Assesses the robustness of validated samples across multiple scheduling algorithms.

    \item \textbf{Outlier Rejection:}
    \begin{equation}
        T(y_i, f) =
        \begin{cases}
            \text{True}, & \text{if } y_i \text{satisfies CFD-validated constraints } f\\
            \text{False}, & \text{otherwise}
        \end{cases}
    \end{equation}
    \begin{equation}
        D(y_i) = \text{ExpertReview}(y_i)
    \end{equation}
    Eliminates unrealistic or thermodynamically infeasible samples.

    \item \textbf{Acceptance Criteria:}
    \begin{equation}
        Acceptable(y_i) = V(y_i) \land C(y_i, S) \land \neg D(y_i)
    \end{equation}
    A data point $y_i$ is accepted if it meets CFD constraints, passes cross-validation, and is not flagged as an outlier.
\end{itemize}

Integrating CFD-based energy constraints within the VAE’s objective function ensures that synthetic samples adhere to real-world power–temperature dynamics while maintaining diversity. This physically grounded validation process enhances the reliability of generated workload configurations, enabling energy-efficient workflow optimization without sacrificing performance in high-performance computing environments.

Algorithms 1 explains the proposed scheduling system, which employs a Physics-Informed Variational Autoencoder (PI-VAE) to generate thermodynamically valid synthetic workload configurations for energy-aware optimization. The model is trained on a real dataset $D$, which includes workflow execution traces with features such as CPU frequency, turnaround time (TAT), energy consumption, and scheduler identifiers. The VAE consists of an encoder network $q_\phi(z|x)$ that maps input data $x$ into a latent space $z$, and a decoder $p_\theta(x|z)$ that reconstructs input-like samples from the latent variables. The training process minimizes a composite loss function comprising three terms: (1) the reconstruction loss, which ensures fidelity to the original data; (2) the Kullback–Leibler (KL) divergence, which regularizes the latent space to follow a standard normal distribution; and (3) a physics-informed penalty term, $\mathcal{L}_{\text{CFD}}$, that incorporates thermal constraints derived from Computational Fluid Dynamics (CFD) simulations.

During training, each mini-batch is encoded into latent parameters $(\mu, \sigma)$, from which latent vectors $z$ are sampled using the reparameterization trick. The decoder then reconstructs each sample $\hat{x}$, and all three loss components are computed and backpropagated to update the encoder and decoder weights. The CFD penalty is evaluated by simulating the decoded configurations and penalizing outputs that exceed predefined power-temperature thresholds, ensuring thermodynamic realism in the learned representations.

Once the PI-VAE is trained, it is used to generate a synthetic dataset $S$ by sampling new latent vectors $z$ and decoding them into configuration samples $\hat{x}$. Each synthetic sample is evaluated against the CFD constraints, and only those satisfying thermal feasibility are retained. The accepted synthetic configurations are then used to simulate workflow executions under various scheduling strategies (e.g., FCFS, LAS, LASP, SAS). For each configuration-scheduler pair, the system computes energy consumption and TAT, forming a set of performance points in a multi-objective space. Finally, efficiency criteria are applied to identify optimal scheduling decisions that provide the best trade-offs between energy usage and execution time. This process enables a principled, data-efficient, and physics-constrained exploration of scheduling options, bridging machine learning with domain-specific energy modeling.



\begin{algorithm}
\caption{Synthetic Data Generation for Energy-Aware Scheduling Using a CFD-PIVAE}
\begin{algorithmic}[1]
\Require $D$: Real dataset of workflow traces $\{x_i\}$ with features: CPU frequency, energy consumption, turnaround time, scheduler ID
\Require $C$: CFD-derived thermal constraints (power-temperature mappings)
\Ensure $S$: Synthetic dataset of thermodynamically valid workflow configurations

\State Define VAE architecture: encoder $q_\phi(z|x)$, decoder $p_\theta(x|z)$ with latent dimension $d$
\State Define composite loss function: $\mathcal{L} = \mathcal{L}_{\text{recon}} + \beta \mathcal{L}_{\text{KL}} + \gamma \mathcal{L}_{\text{CFD}}$

\For{each mini-batch $\{x_i\} \subset D$}
    \State Encode input: $\mu, \sigma = \text{Encoder}(x_i)$
    \State Sample latent vector: $z_i \sim \mathcal{N}(\mu, \sigma^2)$ using reparameterization trick
    \State Decode latent vector: $\hat{x}_i = \text{Decoder}(z_i)$
    \State Compute reconstruction loss: $\mathcal{L}_{\text{recon}} = \|x_i - \hat{x}_i\|^2$
    \State Compute KL divergence: $\mathcal{L}_{\text{KL}} = D_{KL}(q_\phi(z|x_i)\,\|\,\mathcal{N}(0,I))$
    \State Apply CFD constraint: evaluate $\hat{x}_i$ via thermal simulation, compute $\mathcal{L}_{\text{CFD}}$ as penalty if thermal thresholds violated
    \State Backpropagate total loss $\mathcal{L}$ and update $\phi$, $\theta$
\EndFor

\State Initialize empty synthetic dataset $S \leftarrow \{\}$
\For{each sample $z_j \sim \mathcal{N}(0, I)$}
    \State Generate configuration $\hat{x}_j = \text{Decoder}(z_j)$
    \State Evaluate CFD consistency using $C$: if $\hat{x}_j$ violates thermal limits, discard
    \State Add accepted $\hat{x}_j$ to $S$
\EndFor

\State Use $S$ to simulate scheduling outcomes under multiple strategies (FCFS, LAS, LASP, SAS)
\State Compute energy and TAT trade-offs for each configuration
\State Identify optimal scheduling policy per workflow using efficiency criteria
\end{algorithmic}
\end{algorithm}

\subsubsection{Error Handling}
To avoid the inclusion of unrealistic configurations in the synthetic data generated by the Variational Autoencoder (VAE), we restricted the model to specific operational scenarios. These scenarios included adjustments of 5\%, 10\%, 15\, and 20\% reductions in CPU operational configurations. This constraint was imposed to prevent extreme modifications, such as reductions of only 1\% or over 50\%which could severely disrupt the system's overall performance. Extreme decreases in operational configurations, particularly those less than 5\%, show minimal impact on both energy savings and turnaround time reduction. Conversely, reductions exceeding 20\% enhance energy efficiency but at the cost of significantly prolonging turnaround times, thus negatively affecting system performance. This targeted ensures that the synthetic data remains realistic and applicable for optimizing energy efficient scheduling within computational workflows.

Given the operational configurations $\mathcal{C} = \{5\%, 10\%, 15\%, 20\%\}$, we define the model for generating synthetic data via a Variational Autoencoder (VAE) and its validation through predefined acceptable operational limits.

\begin{algorithm}
\caption{Error Handling for PI-VAE Generated Configurations}
\begin{algorithmic}[1]
\Require Set of operational reductions $X = \{5\%, 10\%, 15\%, 20\%\}$
\Require Thermal constraint function $\mathcal{C}(x)$ from CFD model
\Ensure Synthetic data that reflects thermodynamically feasible system performance

\State $x \gets$ Execute Algorithm 1 \Comment{Obtain real configuration samples}
\For{each $x \in X$} 
    \State $Y(x) \gets \text{PI-VAE}(x)$ \Comment{Generate synthetic data for configuration $x$}
    \State Evaluate $f(x)$ \Comment{Measure predicted performance (e.g., energy, TAT)}
    
    \If{$\neg \mathcal{C}(Y(x))$}
        \State \textbf{continue} \Comment{Reject configuration violating thermal limits}
    \EndIf

    \If{$x < 5\%$ or $x > 20\%$}
        \State \textbf{continue} \Comment{Skip out-of-scope configurations}
    \EndIf

    \State \textbf{record} $Y(x)$ and $f(x)$ \Comment{Store valid synthetic outputs}
\EndFor
\State \textbf{return} All valid $Y(x)$ and associated evaluations $f(x)$
\end{algorithmic}
\end{algorithm}

In algorithm 2, the function $\text{PIVAE}(x)$ used the generated synthetic data $S$ in algorithm 1 based on input configuration $x$. The feasibility of each generated configuration is evaluated by the function $D(x)$, which determines if a configuration meets the system's operational criteria as shown in equation 30, where $\mathcal{X}$ represents the input feature space comprising all workflow–scheduler configurations used for model training.

\begin{equation}
Y(x) = \text{PIVAE}(x), \quad \forall x \in \mathcal{X} \end{equation}
\begin{equation}
D(x)  = 
  \begin{cases} 
    \text{True} & \text{if } 5\% \leq x \leq 20\% \\
    \text{False} & \text{otherwise}
  \end{cases}
\end{equation}

The error handling  is modeled as a constraint on the acceptable configurations. The synthetic data must not only simulate realistic scenarios but also conform to specified performance criteria related to energy efficiency and turnaround time (TAT). The acceptable configurations are defined by:

\begin{equation}
\text{Acceptable}(y_i) = 
  \begin{cases} 
    \text{True} & \text{if } D(x_i) = \text{True} \\
    \text{False} & \text{otherwise}
  \end{cases}
\end{equation}

Where $y_i$ is a data instance generated from configuration $x_i$.

The objective is to maximize the utility of synthetic data $S$ while ensuring that all data instances $y_i \in S$ adhere to realistic and permissible operational configurations as defined by $D(x)$. 

\subsection{Validation of Synthetic Data and Uncertainty Quantification}

Various studies have utilized bootstrap resampling for uncertainty quantification due to its effectiveness in estimating the variability of statistical estimates \cite{palmer2022calibration} \cite{endo2015confidence}. 
We employ a bootstrap resampling approach on the difference in means between real measurements and PIVAE-generated synthetic data. This gives a non-parametric estimate of variability without assuming any particular distribution of the underlying data, making it well-suited for heterogeneous HPC workloads.

Given two samples, $X_{\text{real}}$ and $X_{\text{PIVAE}}$, representing metrics such as total energy consumption or turnaround time (TAT), the difference in means is defined as:
\begin{equation}
    \Delta = \bar{X}_{\text{real}} - \bar{X}_{\text{PIVAE}}.
\end{equation}

Bootstrap resampling, as shown in algorithm 3, is performed by repeatedly drawing samples with replacement from $X_{\text{real}}$ and $X_{\text{PIVAE}}$ to generate new empirical distributions of $\bar{X}_{\text{real}}$ and $\bar{X}_{\text{PIVAE}}$. For each iteration, we compute $\Delta_b$, yielding a bootstrap distribution $\{\Delta_1, \Delta_2, ..., \Delta_B\}$ over $B$ trials. This allows estimation of a confidence interval (CI) for $\Delta$:
\begin{equation}
    CI_{1-\alpha} = \left[ Q_{\alpha/2}(\Delta), Q_{1-\alpha/2}(\Delta) \right],
\end{equation}
where $Q_p$ denotes the $p$-th quantile of the bootstrap distribution. 

The bootstrap distribution also supports hypothesis testing. Specifically, a two-sided $p$-value for the null hypothesis $H_0: \Delta = 0$ is estimated from the fraction of resampled statistics on either side of zero. This enables us to assess whether the observed differences between real and synthetic (CFD-informed PIVAE) results are statistically significant.

In the context of our system, this procedure provides a principled way to quantify the reliability of PIVAE-generated energy and performance estimates. By applying bootstrap resampling to metrics across different workflows and schedulers, we capture uncertainty due to limited samples and workload variability, ensuring that the synthetic data maintains not only thermodynamic consistency but also statistical credibility.

\begin{algorithm}
\caption{Uncertainty Quantification using Bootstrap Resampling}
\label{alg:bootstrap}
\begin{algorithmic}[1]
\State \textbf{Input:} Real data $D$, PIVAE-generated data $G$, number of bootstrap samples $B$, confidence level $CL$
\State \textbf{Output:} Confidence interval for the difference in means between $D$ and $G$

\State Normalize the real data $D$ and generated data $G$ 
\State \quad Compute mean and standard deviation of $D$: $\text{mean\_D}$, $\text{std\_D}$
\State \quad Normalize $D$: $R_{\text{normalized}} = \frac{D - \text{mean\_D}}{\text{std\_D}}$
\State \quad Normalize $G$: $G_{\text{normalized}} = \frac{G - \text{mean\_D}}{\text{std\_D}}$

\State Initialize an array to store bootstrap differences: $\text{bootstrap\_diffs}[B]$
\For{each bootstrap sample $i$ from $1$ to $B$}
    \State Resample with replacement from $R_{\text{normalized}}$: $\text{bootstrap\_D} = \text{sample\_with\_replacement}(D_{\text{normalized}})$
    \State Resample with replacement from $G_{\text{normalized}}$: $\text{bootstrap\_G} = \text{sample\_with\_replacement}(G_{\text{normalized}})$
    \State Compute the mean of the bootstrap samples: $\text{mean\_bootstrap\_D} = \text{mean}(\text{bootstrap\_D})$, $\text{mean\_bootstrap\_G} = \text{mean}(\text{bootstrap\_G})$
    \State Compute the difference in means: $\text{diff} = \text{mean\_bootstrap\_D} - \text{mean\_bootstrap\_G}$
    \State Store the difference in $\text{bootstrap\_diffs}[i]$
\EndFor

\State Compute the percentiles for the confidence interval:
\State \quad $\text{lower\_percentile} = (1 - CL) / 2 \times 100$
\State \quad $\text{upper\_percentile} = (1 + CL) / 2 \times 100$

\State Compute the confidence interval for the difference in means:
\State \quad $\text{lower\_bound} = \text{percentile}(\text{bootstrap\_diffs}, \text{lower\_\%ile})$
\State \quad $\text{upper\_bound} = \text{percentile}(\text{bootstrap\_diffs}, \text{upper\_\%ile})$

\State Output the confidence interval: $[\text{lower\_bound}, \text{upper\_bound}]$

\end{algorithmic}
\end{algorithm}

\subsection{Speculative Scheduling: Time and Energy Sweet Spot}
To select an optimal operational configuration and scheduling technique for a specific workflow that minimizes energy consumption and TAT, we define an objective function \( O(F, S) \) that balances energy \( E \) and turnaround time \( T \) for operational configuration \( F \) and scheduling technique \( S \). Given a synthetic dataset from PIVAE, we apply:

\begin{equation}
\min_{F, S} O(F, S) = \alpha E(F, S) + \beta T(F, S)
\end{equation}

where \( \alpha \) and \( \beta \) are weighting factors balancing the importance of energy savings versus TAT.

\begin{algorithm}
\caption{optimisation of Energy Consumption for Workflows}
\begin{algorithmic}[1]
\Require $Data$: Set of workflows $\{WF_1, WF_2, WF_3, WF_4, WF_5\}$
\Require $SchedulingTechniques$: $\{FCFS, LAS, LASP, LYNX, SAS\}$
\Require $FrequencyAdjustments$: $\{5\%,10\%, 15\%, 20\% \}$
\State Initialize evaluation metrics: Total Energy Consumption ($E_{total}$), Turnaround Time ($TAT$)

\For{each $WF$ in $Data$}
    \For{each $schedTech$ in $SchedulingTechniques$}
        \For{each $freqAdj$ in $FrequencyAdjustments$}
            \State Apply $schedTech$ to schedule workflows in $WF$
            \State Adjust operational configuration according to $freqAdj$
            \State Calculate $TAT$ for $WF$ with current $schedTech$ and $freqAdj$
            \State Measure CPU $PowerUsage$ during $WF$ execution
            \State Compute $E_{total}$ using $PowerUsage$ and $TAT$
            \State Store $E_{total}$ and $TAT$ for analysis
        \EndFor
    \EndFor
\EndFor

\State Analyze the impact of $FrequencyAdjustments$ on $E_{total}$ and $TAT$ across all $WF$s
\State Determine optimal $freqAdj$ for each $WF$ and $schedTech$ combination minimizing $E_{total}$ while maintaining acceptable $TAT$
\end{algorithmic}
\end{algorithm}

Algorithm 4 serves as the decision-making core of the proposed CFD-PIVAE system, linking physics-informed modeling with scheduling. For each workflow, multiple schedulers (FCFS, LAS, LASP, LYNX, SAS, OM-FNN) are applied, while processor frequency is adjusted in steps (5\%, 10\%, 15\%, 20\%) to emulate Dynamic Voltage and Frequency Scaling (DVFS). Turnaround time (TAT) and total energy consumption ($E_{total}$) are then evaluated, where CFD-derived thermal profiles provide temperature-dependent power dissipation and PIVAE generated synthetic data extend the range of workload scenarios. Iterating across all workflow scheduler frequency combinations produces an energy performance landscape from which optimal operating points are identified, ensuring energy efficiency without unacceptable increases in TAT.

\subsection{Computational Cost of CFD-PIVAE}

The computational cost of the proposed CFD-PIVAE framework consists of two stages: CFD-based data generation and VAE training. Let $N$ be the number of workflow instances, $F$ the CFD feature dimension, $L$ the number of neural layers, and $d$ the latent-space size.

\textbf{CFD Simulation:} Each CFD run has complexity $O(N \cdot F)$ but is executed once during data preparation. The resulting thermal parameters are reused throughout model training.

\textbf{VAE Training:} Each epoch involves forward and backward passes through $L$ layers, requiring $O(N \cdot L \cdot d)$ operations. With mini-batches of size $B$, the effective per-epoch cost becomes $O((N/B) \cdot L \cdot d)$, which scales linearly with data size.

\textbf{Overall Complexity:}
\begin{equation}
\mathcal{O}(N \cdot F) + \mathcal{O}((N/B) \cdot L \cdot d) \approx O(N)
\end{equation}
for fixed $F, L,$ and $d$. 

Hence, CFD preprocessing adds negligible amortized overhead, and the framework maintains near-linear scalability.

\section{Experimental Environment}

This work used the DiRAC Data Intensive service (DIaL) at the University of Leicester, managed by the University of Leicester Research Computing Service on behalf of the STFC DiRAC HPC Facility (www.dirac.ac.uk). The DiRAC service at Leicester was funded by BEIS, UKRI and STFC capital funding and STFC operations grants. DiRAC is part of the UKRI Digital Research Infrastructure.
The infrastructure comprised dual Intel Xeon Platinum 8280 processors (56 cores, 112 threads per node), 1.5 TB DDR4 ECC RAM, and 10 TB NVMe SSD storage, managed by a Supermicro H11DSi-NT Dual Socket Motherboard with redundant 1400W power supplies. All nodes were interconnected via a 200~Gbps HDR InfiniBand network and shared a 4~PB Lustre parallel file system for high-throughput I/O. Power telemetry was captured at 100 ms intervals using Intel RAPL and IPMI interfaces, while thermal metrics were continuously monitored via LM-Sensors. The software environment included Ubuntu 20.04 LTS for orchestration and CentOS 8 for executing distributed workflow simulations. Resource isolation and fine grained energy profiling were achieved using the Proxmox virtualization platform. To assess the thermal effects of our proposed scheduler, we monitored component-level temperatures (CPU cores and memory DIMMs) via IPMI/BMC sensors. Heat maps were generated before and after applying the CFD-PIVAE based optimization, capturing spatial thermal distributions. 

All scheduling strategies were implemented as modular plug-ins within the DiRAC workflow orchestration layer, using Python bindings over the Slurm API to manage task queues, CPU frequency states, and runtime priorities under identical resource conditions. To evaluate the proposed scheduling system, over 4000 real workflow executions were performed across five representative scientific workflows, under varying CPU frequencies and scheduling strategies. These real traces were used to train the PIVAE model, forming the empirical basis for learning energy performance trade-offs. The synthetic operational scenarios generated by the PIVAE were validated through uncertainty quantification using bootstrap resampling, and the optimal configurations it predicted were re-executed on the same cluster to measure actual energy savings and turnaround time improvements. This real world feedback loop ensures that our model is grounded in physical infrastructure constraints and generalizes beyond theoretical assumptions. Figure 3 illustrates the full system setup.

Table 1 categorizes a series of computational workflows, identified as WF-1 through WF-5, by their level of resource utilization in a high performance computing environment. Each workflow contains various High Energy Physics (HEP) data analysis tasks. 

\begin{itemize}
    \item WF-1 (Event Reconstruction): Classified as a mid to low-level resource utilizer, this workflow involves reconstructing events from raw data collected during experiments. Due to the varying complexity and computational demands associated with different reconstruction tasks, it falls within a broader classification range in terms of resource consumption.
    \item WF-2 (Identifying Particle Trajectories): This workflow requires a moderate level of resources (classified as mid) to track particles through detectors. It's essential for understanding particle interactions but is less demanding than workflows requiring pattern recognition or anomaly detection.
    \item WF-3 (Identifying Collision Points): Also classified as mid-level in resource utilization, this workflow deals with pinpointing the locations where particles collide, which is fundamental in experiments that involve particle accelerators.
    \item WF-4 (Pattern Recognition): This is a high-level resource utilization workflow due to the complexity of identifying patterns from large datasets. It is crucial for detecting regularities or structures in the data, which may signify underlying physical phenomena.
    \item WF-5 (Anomaly Detection): Similar to pattern recognition, anomaly detection is categorized as a high resource utilization workflow. It focuses on identifying data points that deviate from the expected norm, which can be indicative of new or rare events that require further investigation.
    
\end{itemize}

\begin{table}[]
\caption{Resource Utilisation of Different Workflows\cite{alice2015technical}}
\begin{tabular}{|l|l|l|}

\hline
\textbf{Workflow} & \textbf{Workflow Description}     & \textbf{\begin{tabular}[c]{@{}l@{}}Resource \\ Utilization \\ Level\end{tabular}} \\ \hline
WF-1              & Event Reconstruction              & Mid/Low                                                                           \\ \hline
WF-2              & Identifying Particle Trajectories & Mid                                                                               \\ \hline
WF-3              & Identifying Collision Points      & Mid                                                                               \\ \hline
WF-4              & Pattern Recognition               & High                                                                              \\ \hline
WF-5              & Anomaly Detection                 & High                                                                              \\ \hline
\end{tabular}
\end{table}


Each workflow  was parameterized by 500-900 tasks per instance, as table 2 shows the complete experimental matrix.  Energy consumption was calculated as $E = \sum (P_{CPU} \cdot t_{exec})$, where $P_{CPU}$ denotes CPU power (W) and $t_{exec}$ is task execution time.

To address data scarcity, a Physics-Informed Variational Autoencoder (PI-VAE) was trained on around 4000 real workflow executions. The model architecture included 6-layer encoder/decoder networks with latent dimensions constrained by ANSYS Fluent CFD simulations of heat dissipation. These CFD-derived thermal profiles were integrated into the VAE's loss function as regularization terms, enforcing thermodynamic plausibility in synthetic data generation. The total loss function combined reconstruction error ($\mathcal{L}_{rec}$), KL divergence ($\mathcal{L}_{KL}$) and CFD-based energy constraints ($\mathcal{L}_{CFD}$):
$$
\mathcal{L} = \mathcal{L}_{rec} + \beta \mathcal{L}_{KL} + \gamma \mathcal{L}_{CFD},
$$
where $\beta$ and $\gamma$ balanced latent space regularization and physical fidelity.

\begin{figure}[h!]
\centering
\includegraphics[width=0.5\textwidth]{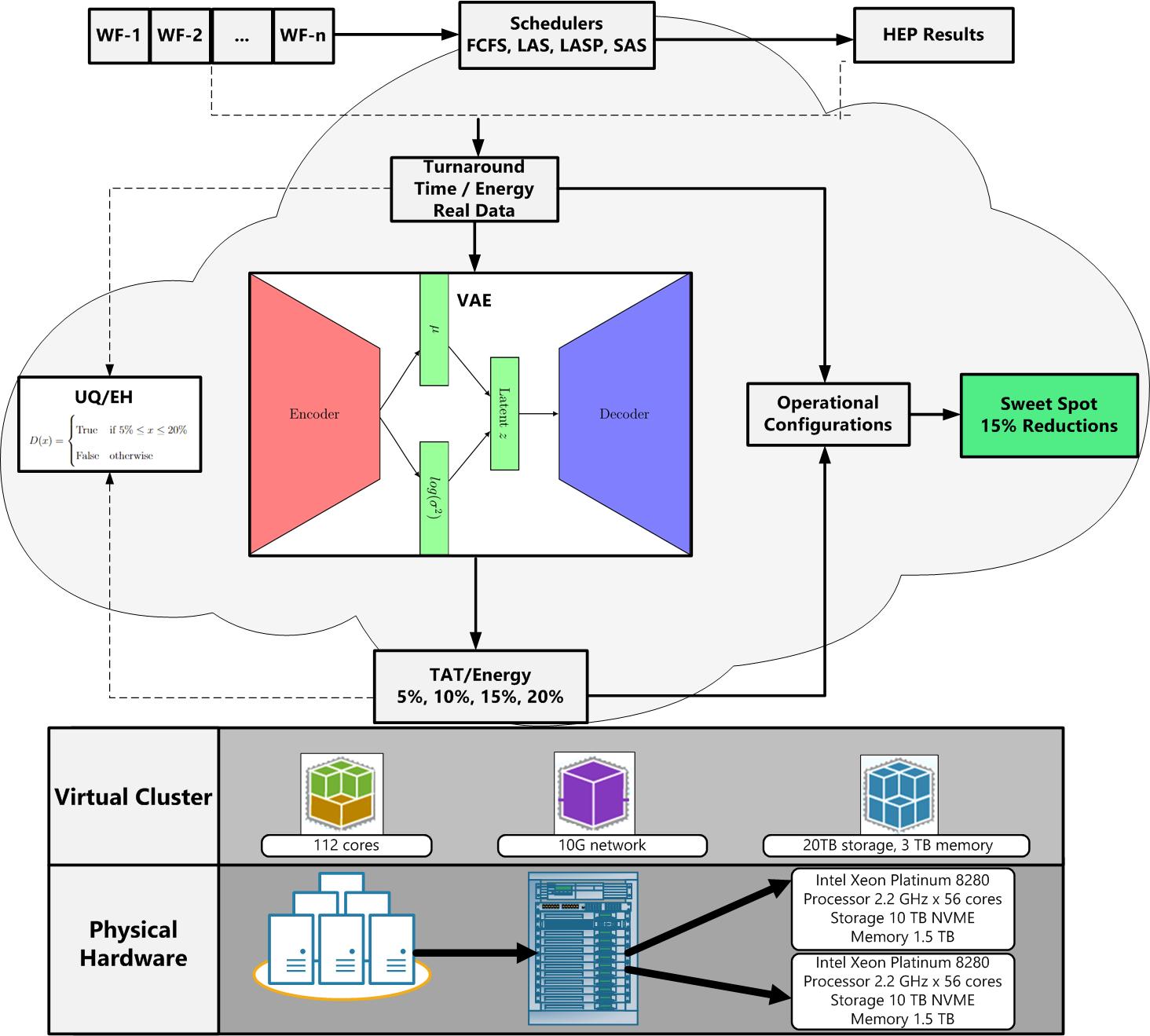}
\caption{Experimental Environment for Energy Aware Scheduling}
\end{figure}


\begin{table}[]
\caption{Experimental matrix}
\begin{tabular}{|l|l|}
\hline
\textbf{Parameters}                                                          & \textbf{Description}                                                                                                                                           \\ \hline
Scheduling                                                                   & FCFS, LAS, LASP, LYNX and SAS                                                                                                                                  \\ \hline
WF Description                                                               & WF-1, WF-2, WF-3, WF-4 and WF-5                                                                                                                                \\ \hline
\begin{tabular}[c]{@{}l@{}}No of Tasks in \\ WF\end{tabular}                 & \begin{tabular}[c]{@{}l@{}}WF-1 = 500, WF-2 = 600, WF-3 =750 ,\\  WF-4=800 and WF-5= 900\end{tabular}                                                          \\ \hline
CPU                                                                          & Operational speed of the CPU: 2.1 GHz                                                                                                                          \\ \hline
\begin{tabular}[c]{@{}l@{}}Real CPU \\ Frequency \\ Utilization\end{tabular} & \begin{tabular}[c]{@{}l@{}}80\% of the total CPU during task \\execution, offering a precise measure \\ of CPU engagement.\end{tabular} \\ \hline
TAT in ms                                                                    & \begin{tabular}[c]{@{}l@{}}The total time taken from the start to \\ the completion of all tasks.\end{tabular}                                                 \\ \hline
\begin{tabular}[c]{@{}l@{}}CPU Power \\ Consumption\end{tabular}             & \begin{tabular}[c]{@{}l@{}}The total power consumed by the CPU\\  during the execution of tasks\end{tabular}                                                   \\ \hline
\begin{tabular}[c]{@{}l@{}}Energy\\ Consumption\\  in (KW/H)\end{tabular}    & \begin{tabular}[c]{@{}l@{}}The cumulative energy used by the CPU \\ during task processing, expressed in (KW/H)\end{tabular}                                   \\ \hline
\end{tabular}
\end{table}


\section{Results and discussion}


The primary objective of the experiments was to evaluate the performance of the proposed CFD-PIVAE framework in optimizing energy consumption and turnaround time (TAT) across heterogeneous HPC workflows. The analysis focused on quantifying the efficiency of various scheduling strategies—FCFS, LAS, LASP, LYNX, and SAS—when applied to five representative scientific workflows (WF-1 to WF-5) executed on the DiRAC DIaL system. Each configuration was evaluated using CFD-informed thermal and energy measurements to assess the trade-off between reduced energy consumption and sustained computational performance.

The results section begins by comparing these scheduling techniques using real data, highlighting the trade-offs between execution time and energy consumption. This is followed by an evaluation of the proposed ML-based approach, which uses a CFD-PIVAE to generate synthetic workflow data. The impact of these synthetic configurations is analyzed in terms of both energy efficiency and TAT improvement. To validate the effectiveness of the ML-generated solutions, thermal heatmaps of system components are presented, showing before-and-after states with respect to optimization. The section concludes with a detailed explanation of the error handling mechanism for synthetic data generation and an uncertainty quantification strategy, using latent sampling and bootstrap resampling to assess the reliability and robustness of the generated configurations.
\subsection{Comparison of Scheduling Strategies: Turnaround Time and Energy Consumption (Real Data)}

To ensure a fair comparison, all scheduling strategies were implemented within the same workflow execution framework on the DiRAC Data Intensive at Leicester (DIaL) system. The baseline schedulers—FCFS, LAS, LASP, LYNX, and SAS—were developed as plug-ins in the DiRAC orchestration layer using Python bindings over the Slurm API, allowing unified control of job queues, CPU frequency states, and task-level priorities. Each scheduler executed identical workloads and resource constraints so that any observed variations in turnaround time (TAT) and energy consumption were attributable solely to scheduling logic. The proposed CFD-PIVAE scheduler was integrated into this environment to dynamically adjust task allocation and CPU power caps via the RAPL interface at runtime. Every configuration was executed five times, and the reported results represent mean values with deviations below~3\%.
\begin{table}[]
\centering
\caption{Task Scheduling and Energy Calculation Parameters and Metrics}
\begin{tabular}{lllllll}
\hline
\textbf{\begin{tabular}[c]{@{}l@{}}Sched\\ technique\end{tabular}} & 
\textbf{\begin{tabular}[c]{@{}l@{}}Work\\ load\end{tabular}} & 
\textbf{Tasks} & 
\textbf{\begin{tabular}[c]{@{}l@{}}TAT\\ (ms)\end{tabular}} & 
\textbf{\begin{tabular}[c]{@{}l@{}}Power\\ Consump\\ (W)\end{tabular}} & 
\textbf{\begin{tabular}[c]{@{}l@{}}Energy\\ Consump\\ (kW/h)\end{tabular}} \\ 
\hline
FCFS & WF-1 & 500 & 505.07 & 890.94 & 12.50 \\
FCFS & WF-2 & 600 & 620.84 & 1095.16 & 18.89 \\
FCFS & WF-3 & 750 & 763.67 & 1347.11 & 28.58 \\
FCFS & WF-4 & 800 & 805.31 & 1420.56 & 31.78 \\
FCFS & WF-5 & 900 & 927.73 & 1636.52 & 42.17 \\
LAS  & WF-1 & 500 & 461.64 & 1473.54 & 18.90 \\
LAS  & WF-2 & 600 & 539.05 & 1720.63 & 25.76 \\
LAS  & WF-3 & 750 & 669.88 & 2138.25 & 39.79 \\
LAS  & WF-4 & 800 & 726.16 & 2317.91 & 46.76 \\
LAS  & WF-5 & 900 & 800.15 & 2554.08 & 56.77 \\
LASP & WF-1 & 500 & 397.66 & 2006.72 & 22.17 \\
LASP & WF-2 & 600 & 452.30 & 2282.43 & 28.68 \\
LASP & WF-3 & 750 & 604.29 & 3049.45 & 51.19 \\
LASP & WF-4 & 800 & 630.73 & 3182.83 & 55.76 \\
LASP & WF-5 & 900 & 684.67 & 3455.05 & 65.71 \\
LYNX & WF-1 & 500 & 406.60 & 2097.92 & 23.70 \\
LYNX & WF-2 & 600 & 462.46 & 2386.17 & 30.65 \\
LYNX & WF-3 & 750 & 617.87 & 3188.05 & 54.72 \\
LYNX & WF-4 & 800 & 644.90 & 3327.48 & 59.61 \\
LYNX & WF-5 & 900 & 700.06 & 3612.08 & 70.24 \\
OM-FNN & WF-1 & 500 & 345.00 & 1830.44 & 17.50 \\
OM-FNN & WF-2 & 600 & 410.00 & 2195.12 & 25.00 \\
OM-FNN & WF-3 & 750 & 530.00 & 2898.11 & 42.80 \\
OM-FNN & WF-4 & 800 & 575.00 & 3090.44 & 49.50 \\
OM-FNN & WF-5 & 900 & 610.00 & 3540.98 & 60.00 \\
SAS  & WF-1 & 500 & 317.41 & 2152.96 & 18.98 \\
SAS  & WF-2 & 600 & 380.89 & 2583.55 & 27.33 \\
SAS  & WF-3 & 750 & 492.72 & 3342.09 & 45.74 \\
SAS  & WF-4 & 800 & 531.47 & 3604.95 & 53.22 \\
SAS  & WF-5 & 900 & 584.62 & 3965.44 & 64.40 \\
\hline
\end{tabular}
\end{table}

Table 3 summarizes the comparative results for turnaround time, CPU power consumption, and total energy usage across all workflows. Among the baseline schedulers, the Speculative-Aware Scheduling (SAS) policy consistently achieved the lowest TAT, with WF-5 completing in 584.6~ms compared to 927.7~ms under FCFS. Although FCFS exhibited the lowest instantaneous power draw (42.2~kWh for WF-5), its longer execution time resulted in inferior overall performance. LAS and LASP improved TAT relative to FCFS (e.g., LASP reduced WF-3 from 763.7~ms to 604.3~ms) but increased total energy consumption due to higher CPU utilization and I/O intensity (e.g., LASP consumed 65.7~kWh versus 42.2~kWh for FCFS).

\begin{figure}[h!]
\centering
\includegraphics[width=0.5\textwidth]{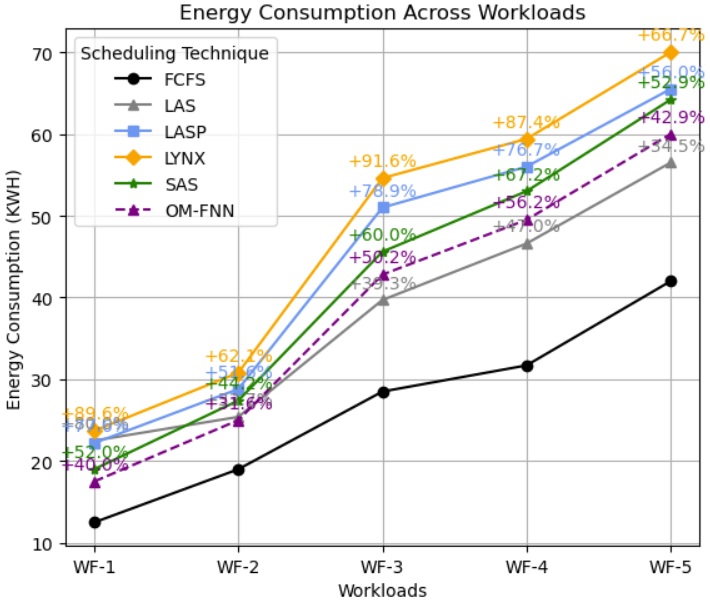}
\caption{Turnaround Time Comparison for All Workflows (Real Data)}
\end{figure}

In general, FCFS provides predictable but slower task completion as workload scale increases, reflecting its lack of adaptivity. LAS, LASP, and SAS leverage locality and speculative execution to achieve faster completion times, yet at differing energy costs. SAS demonstrates the most balanced trade-off: aggressive speculation enables significant TAT reduction while containing the energy overhead through efficient CPU frequency scaling guided by CFD-PIVAE predictions.

A closer inspection of WF-1 illustrates this contrast. Under LAS, tasks completed in approximately 461.6~ms at an average effective CPU frequency of 1.70~GHz (81\% utilization), resulting in a power draw of 1473.5~W and an energy cost of 18.90~kWh. SAS reduced the same workload’s completion time to 317.4~ms by increasing CPU frequency to 2.02~GHz (96\% utilization), drawing 2153~W and consuming 18.98~kWh—slightly higher instantaneous power but markedly improved throughput.

\textbf{Turnaround Time Trends:}
Figure 4 compares TAT across scheduling strategies and workflows. SAS outperformed all others, reducing TAT by up to 37\% relative to FCFS (e.g., WF-5: 927.7~ms~$\rightarrow$~584.6~ms). LASP and LYNX achieved moderate improvements in mid-scale workloads (e.g., WF-3) with 20–21\% reductions, while LAS yielded modest gains of 13–15\%. The widening performance gap for larger workflows indicates that SAS scales more efficiently with workload complexity.

\textbf{Energy Consumption Trends:}
Figure 5 presents the corresponding energy consumption. FCFS recorded the lowest raw energy usage due to minimal scheduling overhead but suffered from high latency. LASP, LYNX, and SAS improved TAT at varying energy costs—LYNX incurred a 91.5\% overhead, LASP 79.1\%, and SAS 60.1\% compared with FCFS for WF-3. Despite this, SAS delivered the most favorable energy–performance ratio, maintaining moderate energy usage while significantly shortening TAT. This trend underscores that as workflows scale, energy efficiency becomes increasingly sensitive to scheduler design; speculative, prediction-driven methods such as SAS and CFD-PIVAE offer the most practical balance for high-throughput HPC environments.

\begin{figure}[]
\centering
\includegraphics[width=0.5\textwidth]{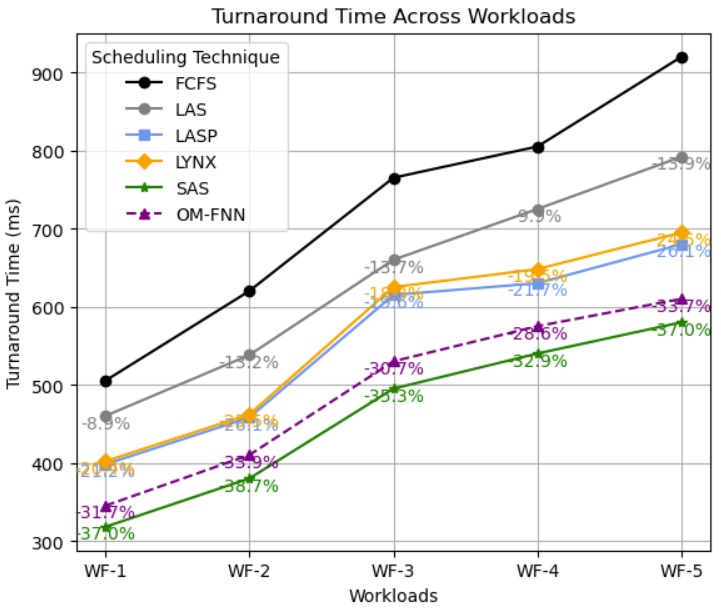}
\caption{Energy Consumption Comparison for all Workflows (Real Data)}
\end{figure}

\subsection{Impact of CFD-PIVAE on Turnaround Time and Energy Consumption}

Table 4 presents a detailed comparison of turnaround time (TAT) and energy consumption for all five scheduling techniques across five workloads (WF-1 to WF-5), using CFD-PI-VAE-generated synthetic data. The baseline configuration (real CPU setting) is evaluated alongside 10\% and 15\% CPU reduction scenarios to examine the trade-offs in performance and efficiency. 

For instance, under FCFS for WF-5, reducing the CPU frequency by 15\% increased TAT from 927.7\,ms to 981.7\,ms (a 5.8\% rise), while reducing energy consumption from 42.17\,kWh to 38.18\,kWh. Similar trends are seen for LAS, where for WF-5, the TAT rose from 800.1\,ms to 840.6\,ms (5.1\%), and energy dropped from 55.7\,kWh to 53.4\,kWh. Notably, SAS demonstrated the most balanced efficiency, where WF-5 under a 15\% CPU reduction yielded only a 5.5\% increase in TAT (from 584.6\,ms to 617.1\,ms), while saving ~2\,kWh (from 58.7 to 56.5\,kWh). 

Across all techniques, energy savings consistently ranged between 3–6\% for 10\% CPU reductions and 5–8\% for 15\% reductions, with minor TAT penalties as shown in the table 4. These results validate that CFD-informed synthetic sampling supports practical, thermodynamically feasible decisions for CPU scaling, offering a viable path for energy-aware scheduling optimization in HPC systems.


\begin{table}[]
\centering
\caption{Task Scheduling and Energy Calculation Parameters and Metrics: (CFD-PIVAE Generated Synthetic Data)}
\begin{tabular}{lllllll}
\hline
\textbf{\begin{tabular}[c]{@{}l@{}}Sched-\\ -uling\\ tech\end{tabular}} & 
\textbf{\begin{tabular}[c]{@{}l@{}}Work \\ flow\end{tabular}} & 
\textbf{\begin{tabular}[c]{@{}l@{}}TAT\\ (ms)\end{tabular}} & 
\textbf{\begin{tabular}[c]{@{}l@{}}TAT\\ (ms)\\ 15\% CPU\end{tabular}} & 
\textbf{\begin{tabular}[c]{@{}l@{}}TAT\\ (ms)\\ 10\% CPU\end{tabular}} & 
\textbf{\begin{tabular}[c]{@{}l@{}}Energy\\ (kW/h)\\ 15\%\end{tabular}} & 
\textbf{\begin{tabular}[c]{@{}l@{}}Energy\\ (kW/h)\\ 10\%\end{tabular}} \\
\hline
FCFS & WF-1 & 505.07 & 535.09 & 526.90 & 11.75 & 12.29 \\
FCFS & WF-2 & 620.84 & 653.11 & 648.69 & 16.34 & 17.15 \\
FCFS & WF-3 & 763.67 & 807.49 & 794.42 & 25.97 & 26.93 \\
FCFS & WF-4 & 805.31 & 850.14 & 840.67 & 30.14 & 31.18 \\
FCFS & WF-5 & 927.73 & 981.74 & 967.84 & 38.18 & 39.67 \\
LAS & WF-1 & 461.64 & 488.34 & 481.02 & 17.80 & 18.47 \\
LAS & WF-2 & 539.05 & 568.67 & 562.62 & 23.44 & 24.40 \\
LAS & WF-3 & 669.88 & 710.03 & 699.61 & 34.60 & 36.06 \\
LAS & WF-4 & 726.16 & 767.83 & 757.34 & 42.20 & 44.27 \\
LAS & WF-5 & 800.15 & 840.65 & 836.06 & 53.41 & 55.70 \\
LASP & WF-1 & 397.66 & 421.36 & 414.59 & 20.15 & 20.85 \\
LASP & WF-2 & 452.30 & 475.38 & 472.26 & 26.91 & 28.09 \\
LASP & WF-3 & 604.29 & 635.01 & 631.06 & 46.31 & 48.30 \\
LASP & WF-4 & 630.73 & 662.48 & 658.77 & 49.02 & 50.78 \\
LASP & WF-5 & 684.67 & 721.11 & 713.09 & 57.56 & 59.93 \\
LYNX & WF-1 & 406.60 & 429.34 & 423.42 & 20.14 & 21.05 \\
LYNX & WF-2 & 462.46 & 489.98 & 481.24 & 27.70 & 28.98 \\
LYNX & WF-3 & 617.87 & 651.55 & 642.79 & 50.89 & 53.40 \\
LYNX & WF-4 & 644.90 & 682.15 & 672.86 & 53.84 & 56.42 \\
LYNX & WF-5 & 700.06 & 741.68 & 729.08 & 61.26 & 63.91 \\
SAS & WF-1 & 317.41 & 333.96 & 330.60 & 17.75 & 18.52 \\
SAS & WF-2 & 380.89 & 402.36 & 396.32 & 24.83 & 25.78 \\
SAS & WF-3 & 492.72 & 520.23 & 512.98 & 40.30 & 41.92 \\
SAS & WF-4 & 531.47 & 560.01 & 554.20 & 48.17 & 50.19 \\
SAS & WF-5 & 584.62 & 617.13 & 609.79 & 56.51 & 58.71 \\
\hline
\end{tabular}
\end{table}

\textbf{Operational Configurations Reduction Impact on Turnaround Time:}
The impact of operational configuration reduction on turnaround time (TAT) was further analyzed under the CFD-PIVAE–guided scheduling framework. Figure 6 illustrates that as operational configurations are reduced, both energy consumption and TAT exhibit a correlated response. For example, in Workflow~5 (WF-5), CFD-PIVAE optimized the SAS scheduler by reducing configuration utilization by 20\%, lowering energy consumption from 62.38~kWh to 54.65~kWh—a 12.4\% saving—while maintaining comparable TAT. These results demonstrate that the proposed model effectively constrains speculative scheduling decisions within thermodynamically feasible limits, achieving tangible energy gains without compromising performance.

\begin{figure}[h!]
\centering
\includegraphics[width=0.5\textwidth]{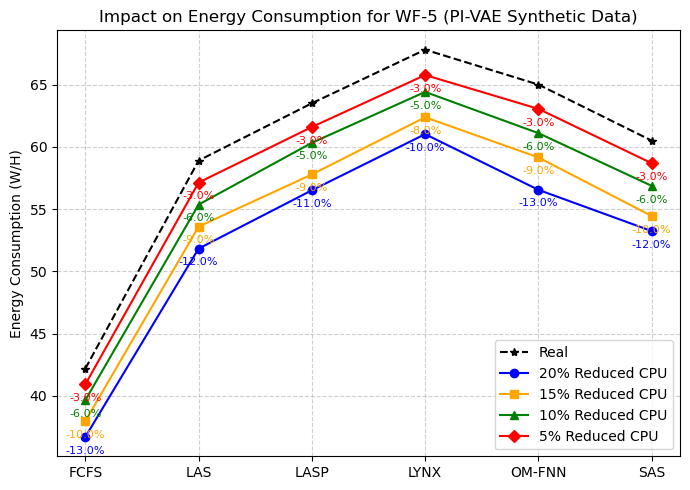}
\caption{CPU Impact on Energy Consumption for WF-5 under Different Scheduling (PIVAE Synthetic Data)}

\end{figure}

\textbf{Operational Configurations Reduction Impact on Heat Signatures}

To evaluate the thermal impact of the proposed energy-aware scheduling system, we monitored component-level temperatures (CPU cores and memory DIMMs) using built-in system management utilities (IPMI/BMC). Heat maps were generated before and after applying the PIVAE-based optimization, with temperature data aggregated and visualized using a gradient scale to capture spatial thermal distributions. As shown in Figure 7, the proposed algorithm consistently reduced thermal output, achieving an average 5–7\% decrease in component temperatures. This reduction validates the efficacy of our approach in lowering energy usage and highlights additional benefits in terms of cooling efficiency and system longevity.

\begin{figure}[h!]
\centering
\includegraphics[width=0.4\textwidth]{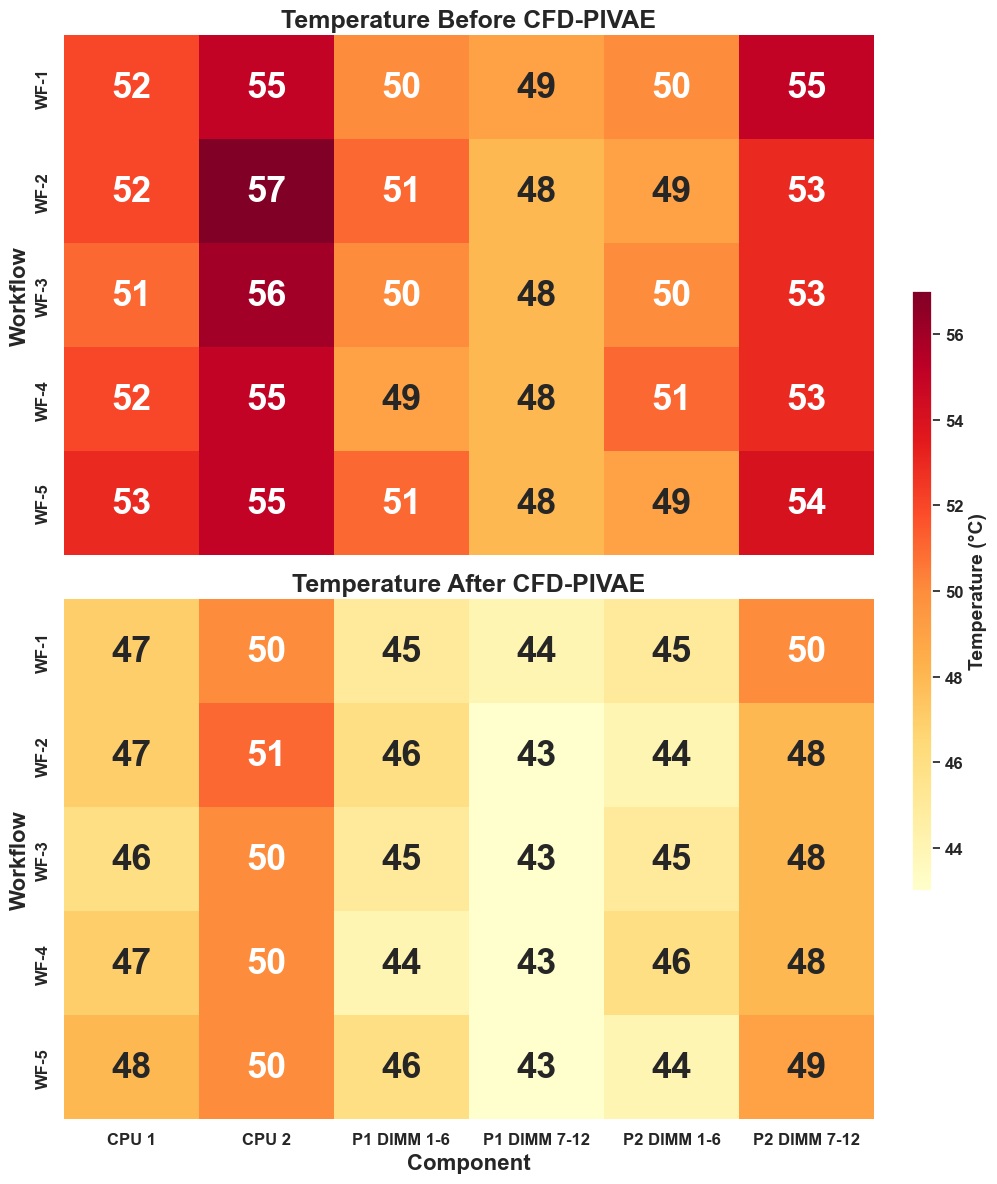}
\caption{Heatmap of Component Temperatures Across Five Workflows Before and After Optimisation}
\end{figure}

\subsection{Uncertainty Quantification using Bootstrap Resampling}

Figure 8 presents the bootstrap uncertainty quantification for the difference in mean energy consumption between real and PIVAE-generated samples. The histogram shows the empirical distribution of $B = 10{,}000$ resampled mean differences, with the 95\% confidence interval (CI) estimated as $[0.715,\, 3.354]$~kWh. The observed mean difference is approximately $2.03$~kWh, indicating that real energy measurements are consistently higher than the synthetic estimates produced by CFD-PIVAE. The bootstrap $p$-value of $0.0022$ provides strong evidence that this difference is statistically significant.

Rather than implying a systematic bias, this offset reflects the expected outcome of optimization: the PIVAE-generated configurations correspond to energy-efficient operating states predicted by the model, whereas the real measurements represent unoptimized, measured workloads. The statistical separation between the two distributions therefore suggests that the model has learned meaningful, physically valid relationships that enable reduction of energy consumption without distorting underlying thermal–performance trends.

While a final validation through controlled re-execution of selected workflows under the model-recommended settings would provide further confirmation, the present bootstrap analysis demonstrates that the CFD-PIVAE system captures consistent, thermodynamically plausible improvements. This reinforces the reliability of its synthetic data for evaluating energy–performance trade-offs in HPC scheduling.

\begin{figure}[h!]
\centering
\includegraphics[width=0.5\textwidth]{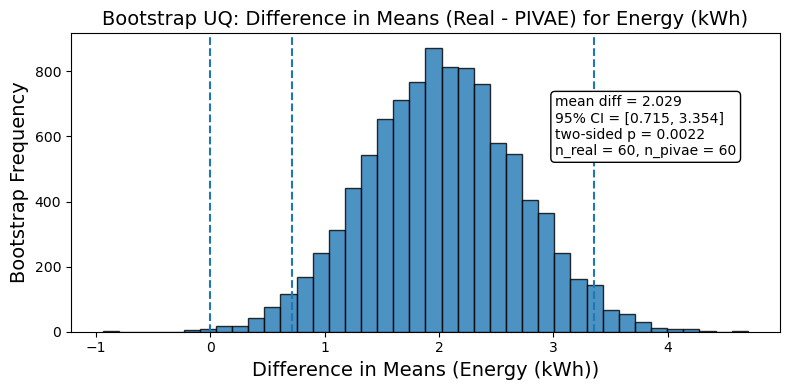}
\caption{Uncertainty Quantification: Bootstrap Resampling Difference in means: Energy Consumption}
\end{figure}

\subsection{Speculative Scheduling: Time and Energy Sweet Spot}


Figure 9 illustrates the relationship between energy consumption and turnaround time (TAT) when operational configurations  performance is decreased using various scheduling techniques. When the CPU's performance is decreased to 20\%, 15\%, 10\%, and 5\% of its normal operating frequency, we notice a rise in TAT (turnaround time) for all scheduling techniques. This is to be expected, as there is less computational power available to handle tasks. The decrease in processing capability results in higher TAT, which range from approximately 10.02\% to 10.86\% when the CPU is reduced by 20\%.


In contrast, there is a proportional decrease in energy consumption, indicating that the system is utilising less power as a result of the diminished performance. Notably, the reduction in energy usage is more significant than the rise in TAT. For instance, in the case of SAS, when the CPU is decreased by 20\%, the energy consumption is reduced by approximately 12.76\%, which surpasses the increase in Turnaround Time (TAT). The observed pattern remains consistent among the other scheduling techniques and configuration reductions, albeit with slight variations in the ratios.

As we decrease the configuration reduction from 20\% to 5\%, the TAT (turnaround time) progressively decreases, indicating diminishing returns in terms of energy savings compared to time performance. For example, when the CPU is reduced to 95\%, the techniques experience an increase in TAT (turnaround time) ranging from approximately 3.05\% to 3.90\%. Additionally, there is a decrease in energy consumption ranging from about 4.14\% to 4.81\%. This suggests that once a certain threshold is reached, the amount of energy saved by decreasing CPU performance results in diminishing returns, while still negatively affecting performance.

The findings indicate a trend: a decrease of 15\% in operational configuration has a disproportionately smaller effect on the turnaround time (TAT) for processing tasks, but it does significantly decrease energy consumption. The TAT increase across various scheduling techniques is only around 5-6\%, which is relatively moderate compared to the energy savings of approximately 9-10\%. This suggests that by reducing the CPU usage to 15\% of its maximum capacity for similar workflow analysis, a significant portion of its processing efficiency can be maintained while achieving noticeable energy savings.

\begin{figure}[h!]
\centering
\includegraphics[width=0.5\textwidth]{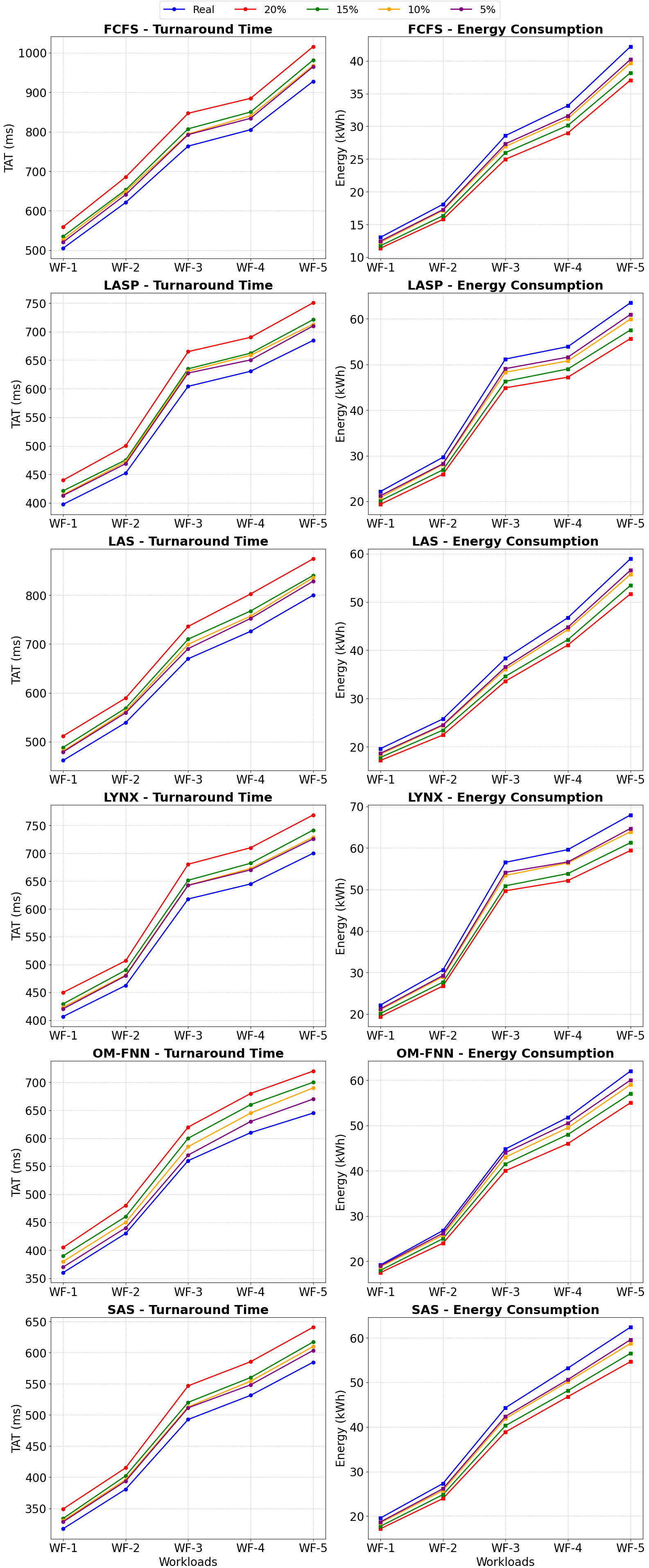}
\caption{Impact of reduced CPU frequency on workflow Turnaround Times under different scheduler configurations.(PIVAE Synthetic Data)}

\end{figure}
Within the context of PIVAE generated synthetic data, achieving a 15\% reduction effectively balances the trade off between computational performance and power efficiency. It maintains the majority of the CPU's computational capacity, allowing tasks to be completed in a reasonable amount of time, while also achieving significant energy savings. 


Based on the synthetic data–driven experiments conducted in this study, a CPU frequency reduction of approximately 15\% was observed to offer the most favorable trade-off between energy consumption and performance degradation for the evaluated workflows and scheduling algorithms. While this observation is specific to the experimental setup and workload characteristics used here, it suggests that moderate frequency scaling can provide meaningful energy savings in similar high-performance computing scenarios without severely impacting turnaround time.

\section{Conclusions and Future Work}

This paper presented a physics-informed, data-driven framework for energy-aware scheduling in High-Performance Computing (HPC) environments, explicitly targeting the joint optimization of energy consumption and turnaround time (TAT). By integrating Computational Fluid Dynamics (CFD) with a Physics-Informed Variational Autoencoder (PI-VAE), the proposed system embeds thermodynamic constraints directly into synthetic data generation, ensuring that predicted scheduling configurations remain both physically plausible and operationally realistic. Unlike purely heuristic or statistical approaches, this integration enables scheduling decisions to be guided by the coupled effects of compute load, energy dissipation, and thermal behavior.

Experimental evaluations across multiple scheduling strategies (FCFS, LAS, LASP, LYNX, SAS, and OM-FNN) demonstrated the effectiveness of the proposed approach in identifying a practical operational sweet spot. Specifically, controlled CPU frequency reductions of approximately 15\% achieved energy savings of up to 10\%, while incurring only a 5–6\% increase in turnaround time. These results highlight that meaningful reductions in energy consumption can be achieved without significantly compromising workflow performance, thereby supporting sustainable HPC operation under realistic performance constraints. Furthermore, bootstrap-based uncertainty quantification validated the statistical consistency of PI-VAE–generated results against real execution data, reinforcing confidence in the robustness and reliability of the proposed framework.

The current implementation primarily targets CPU-intensive scientific workflows, where energy consumption is strongly correlated with computational demand and thermal effects factors that are effectively captured through CFD-based modeling. While workflow scale is varied through task counts, input sizes, and CPU frequency settings to reflect different execution scenarios, this representation does not yet fully encompass memory-intensive or I/O-intensive workflows, whose energy and performance characteristics are influenced by memory bandwidth, data movement, and network behavior. As a result, the present model focuses on representing and optimizing energy–time trade-offs for compute-dominated workloads, rather than all possible HPC workflow classes.

Future work will extend this framework to incorporate memory and I/O behavior, enabling a more comprehensive representation of energy and performance dynamics across diverse workflow categories. Planned extensions include integrating memory bandwidth and I/O models into the PI-VAE architecture, as well as conducting systematic comparative evaluations across CPU-, memory-, and I/O-intensive workloads. These enhancements will further strengthen the proposed framework as a general-purpose, physics-informed solution for energy-efficient and performance-aware scheduling in next-generation HPC systems.


%
\bibliographystyle{IEEEtran}
\bibliography{References}
%

\section*{Author Biographies}
\vspace{-0.5cm}
\begin{IEEEbiography}[{\includegraphics[width=1in,height=1.25in,clip,keepaspectratio]{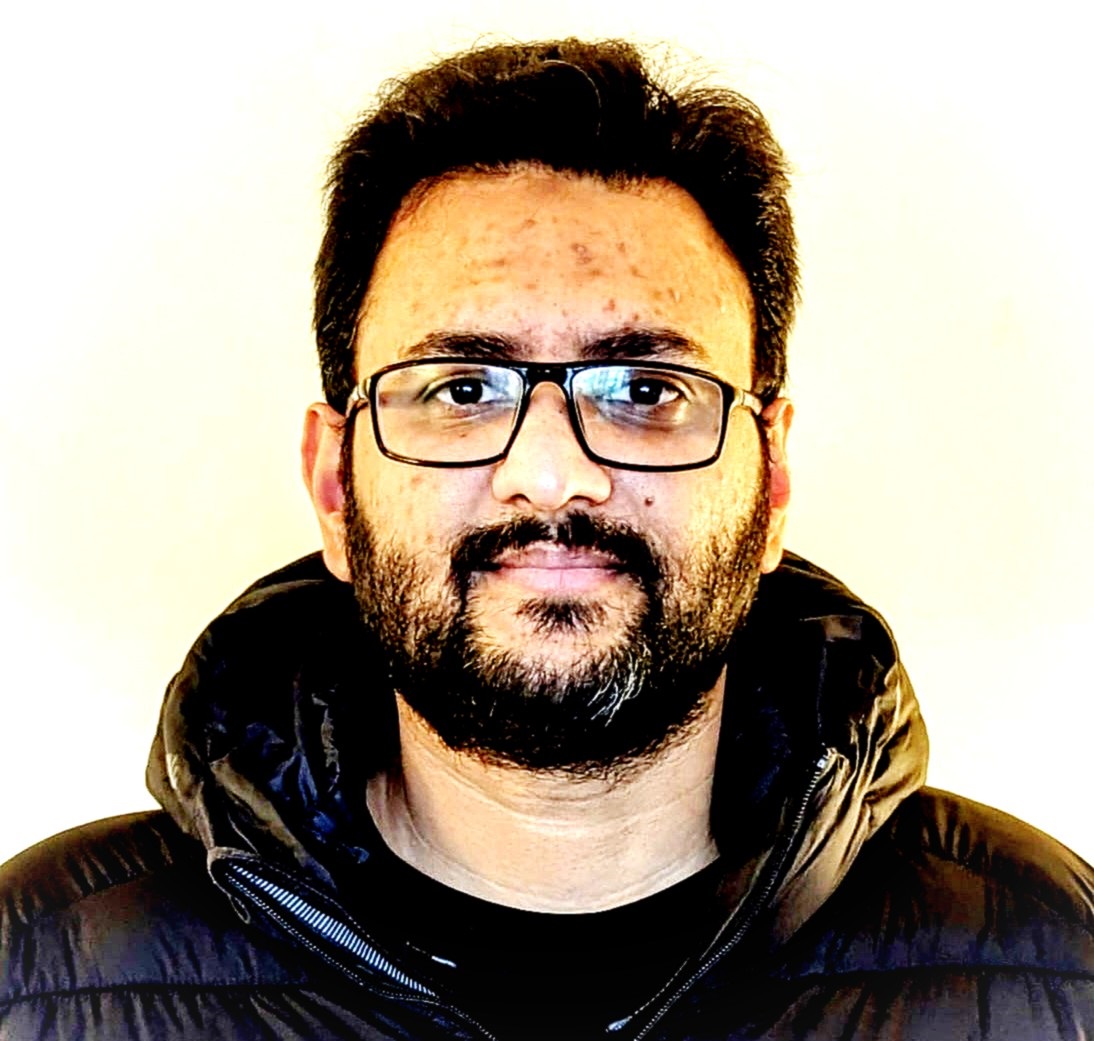}}]{Dr. Ali Zahir}
is a Postdoctoral Fellow at the School of Computing and Mathematical Sciences, University of Leicester, United Kingdom. His research focuses on enhancing data processing workflows in cloud and edge computing environments, with emphasis on High-Energy Physics (HEP). He has contributed to advanced data retrieval algorithms and distributed systems, and has collaborated with CERN on the ALICE project since 2009. Beyond academia, Mr.~Zahir engages in research initiatives in data science and high-performance computing, bridging the gap between theory and implementation in distributed computing. (e-mail: \texttt{ali.zahir4@gmail.com})
\end{IEEEbiography}

\vspace{-0.5cm}

\begin{IEEEbiography}[{\includegraphics[width=1in,height=1.25in,clip,keepaspectratio]{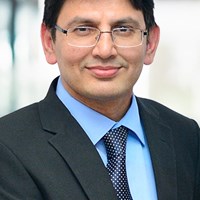}}]{Prof. Ashiq Anjum}
is a Professor of Distributed Systems at the School of Computing and Mathematical Sciences, University of Leicester, where he also serves as the Director of Enterprise and Impact. He is the AI Lead for the £60 million METEOR project at Space Park Leicester. Previously, he was a Professor of Distributed Systems and Director of the Data Science Research Centre at the University of Derby, U.K. His research interests include data-intensive distributed systems, distributed machine learning, self-learning digital twins, and high-performance analytics for streaming data. He has collaborated with Rolls-Royce, BT, and healthcare providers on AI and distributed computing projects. Professor Anjum has an extensive publication record and an H-index of 39. (e-mail: \texttt{aa1180@leicester.ac.uk})
\end{IEEEbiography}

\vspace{-0.5cm}

\begin{IEEEbiography}[{\includegraphics[width=1in,height=1.25in,clip,keepaspectratio]{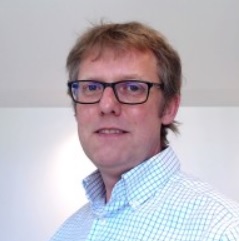}}]{Prof. Mark I. Wilkinson}
is with the Department of Physics and Astronomy, University of Leicester, United Kingdom. He obtained a B.A. and M.Sc. in Theoretical Physics from Trinity College Dublin and a D.Phil. in Theoretical Astrophysics from the University of Oxford under Professor Wyn Evans. His research focuses on the dynamical modelling of dark matter in galaxies, gravitational lensing, and the application of machine learning to astrophysical problems. Prof.~Wilkinson currently serves as Director of the DiRAC High-Performance Computing Facility, leading national efforts in computational astrophysics and large-scale data analysis. (e-mail: \texttt{miw6@leicester.ac.uk})
\end{IEEEbiography}

\vspace{-0.5cm}

\begin{IEEEbiography}[{\includegraphics[width=1in,height=1.25in,clip,keepaspectratio]{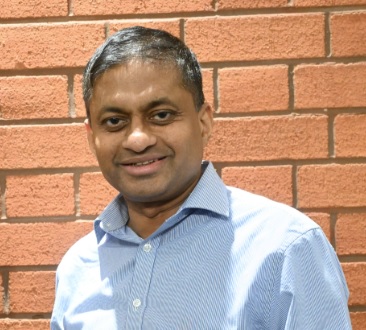}}]{Dr. Jeyan Thiyagalingam} leads the Scientific Machine Learning (SciML) Research Group at the Rutherford Appleton Laboratory, Science and Technology Facilities Council (STFC–UKRI), Harwell, United Kingdom. His group develops and applies machine learning and signal processing techniques to address complex scientific challenges. Prior to joining STFC–RAL, Dr.~Thiyagalingam was an Assistant Professor at the University of Liverpool and held research appointments at the University of Oxford, including a James Martin Fellowship. His interests span scientific machine learning, data-driven modelling, and advanced signal processing. He is a Fellow of the British Computer Society (BCS), a Senior Member of the IEEE, and an Associate Editor for \textit{Patterns}, a Cell Press journal on data science and AI. (e-mail: \texttt{t.jeyan@stfc.ac.uk})
\end{IEEEbiography}

\end{document}